\documentclass[11pt,a4paper]{article}
\usepackage{jheppub}

\title{\boldmath Conical defects and $\mathcal{N}=2$ higher spin holography}

\author{Yasuaki Hikida}

\affiliation{Department of Physics, and Research and Education
Center for Natural Sciences,
Keio University, Hiyoshi, Yokohama 223-8521, Japan}

\emailAdd{hikida@phys-h.keio.ac.jp}

\abstract{
We study conical geometry with the maximal number of fermionic symmetry 
in the higher spin supergravity described by $\text{sl}(N+1|N)  \oplus  \text{sl} (N+1|N)$ 
Chern-Simons gauge theory.
It was proposed that a three dimensional ${\cal N}=2$ higher spin supergravity is
holographically dual to the ${\cal N}=(2,2)$ $\mathbb{CP}^N$ Kazama-Suzuki model.
Based one the duality, we find a map between conical geometries and primary
states in the dual CFT. 
In particular, we construct geometric solutions corresponding
to primary states in the RR-sector. The proposal is checked by the comparison of a few charges and
by the relation between null vectors and higher spin symmetry.
}


\arxivnumber{1212.4124}

\begin{document}

\maketitle
\flushbottom

\section{Introduction}

Higher spin gauge theories include gauge fields with spin two and higher,
and they can be thought as a kind of extensions of gravity theory with
spin two gauge field. They attract a lot of attention since it is believed that 
they are related to the tensionless limit of superstring theory.
Furthermore, higher spin gauge theories on anti-de Sitter (AdS) background are proposed to
be dual to vector-like conformal models with one-less dimensions.
In \cite{Sezgin:2002rt,Klebanov:2002ja} it was proposed that a four dimensional
higher spin gauge theory developed by Vasiliev \cite{Vasiliev:2003ev} is dual to
three dimensional O$(N)$ vector model.
There is also a proposal in lower dimensions that a three dimensional higher
spin gauge theory in \cite{Prokushkin:1998bq} is dual to a large $N$ minimal
model \cite{Gaberdiel:2010pz,Gaberdiel:2012uj}. 
In lower dimensions, there is a possibility that we can understand the duality quite deeply.
This is because three dimensional gravity theory is known to be topological while
two dimensional conformal field theory (CFT) is much restricted due to the large amount of symmetry.
In this paper, we investigate an aspect of ${\cal N}=2$ supersymmetric version
of the duality with lower dimensions in \cite{Creutzig:2011fe}.
Concretely, we study maximally supersymmetric conical defect (surplus) geometry in higher spin supergravity, 
and compare it to the primary states in the dual CFT.

The gravity side of the duality of \cite{Gaberdiel:2010pz} is given by a bosonic truncation
of ${\cal N}=2$ higher spin supergravity proposed by Prokushkin and Vasiliev in 
\cite{Prokushkin:1998bq}. The gravity theory consists of gauge fields with higher 
spins $s=2,3,\ldots$ and massive scalar fields with mass $M^2 = - 1 + \lambda^2$.
The gauge sector can be described by the Chern-Simons theory based on higher spin
algebra hs$[\lambda]$, which can be truncated to sl$(N)$ at $\lambda = \pm N$.
The asymptotic symmetry near the boundary of AdS is found to be 
a large $N$ limit of higher spin $W_N$ algebra called as $W_{\infty}[\lambda]$
\cite{Campoleoni:2010zq,Henneaux:2010xg,Gaberdiel:2011wb,Campoleoni:2011hg,Gaberdiel:2012ku}.
On the other hand, the CFT side is a minimal model with respect to the $W_N$ algebra, 
which can be described by the coset
\begin{align}
 \frac{\text{su}(N)_k \oplus \text{su}(N)_1}{\text{su}(N)_{k+1}} 
 \label{coset}
\end{align}
with the central charge 
\begin{align}
 c = (N-1) \left( 1 - \frac{N(N+1)}{(N+k)(N+k+1)}\right) \, .
\end{align}
Original proposal is that the gravity theory is dual to the 't Hooft limit of the CFT, where
large $N,k$ limit is taken with keeping the 't Hooft parameter
\begin{align}
 \lambda = \frac{N}{k+N}
 \label{thooft}
\end{align} 
finite. Then the parameter is identified with the one in the algebra hs$[\lambda]$ and 
the mass of the dual scalars. There are many works on this duality, and in particular,
the agreement of the spectrum has been shown in \cite{Gaberdiel:2011zw}.
Moreover, holography involving minimal model with so$(N)$ instead of su$(N)$ 
has been proposed in \cite{Ahn:2011pv,Gaberdiel:2011nt} and further refined in 
\cite{Candu:2012ne}. Supersymmetric extensions have been done in 
\cite{Creutzig:2011fe} for ${\cal N}=2$ holography and \cite{Creutzig:2012ar} for 
${\cal N}=1$ holography.

Classical geometry in higher spin gravity has been studied as well.
A higher spin black hole was constructed in \cite{Gutperle:2011kf} (see \cite{Ammon:2012wc} and references therein),
and conical defects are examined in \cite{Castro:2011iw}.
There is a large amount of gauge symmetry in higher spin gravity, and notions like horizon and singularity 
are not gauge invariant. In particular, it was shown that conical defects with the trivial holonomy 
in $\text{sl}(N) \, \oplus \, \text{sl}(N)$ Chern-Simons theory are
mapped by gauge transformation into geometry without any conical singularity. 
It was claimed in \cite{Castro:2011iw} and later refined in \cite{Perlmutter:2012ds}
that the smooth geometry is dual to a primary state in a limit of the $W_N$ minimal model \eqref{coset}.
Other states in the minimal model correspond to perturbative scalar fields in the gravity theory or
their bound states with the conical geometry.
The central charge of the model satisfies $c \leq  N-1$, but the limit is given by an analytic continuation as
$c \to \infty$ with finite $N$. The limit may be called as ``semi-classical'' limit.
A justification of the analytic continuation is given in \cite{Gaberdiel:2012ku}.

The gravity theory for the ${\cal N}=2$ higher spin holography by \cite{Creutzig:2011fe}
is the full ${\cal N}=2$ higher spin supergravity by Prokushkin and Vasiliev \cite{Prokushkin:1998bq}.
This theory includes fermionic higher spin gauge fields in addition to bosonic higher spin gauge fields,
and they are described by the Chern-Simons theory with $\text{shs}[\lambda] \oplus \text{shs} [\lambda]$ superalgebra,
where $\text{shs}[\lambda]$ reduces to  $\text{sl}(N+1|N)$ for $\lambda = N+1$. The theory also includes 
massive scalars and fermions whose masses are organized by the parameter $\lambda$.
The asymptotic symmetry is found to be a large $N$ limit of ${\cal N}=(2,2)$ super $W_{N+1}$ algebra,
which may be called as $SW_{\infty}[\lambda]$ 
\cite{Creutzig:2011fe,Henneaux:2012ny,Hanaki:2012yf,Ahn:2012fz,Candu:2012tr}.
The dual CFT is proposed to be the
$\mathbb{CP}^N$ Kazama-Suzuki model \cite{Kazama:1988qp,Kazama:1988uz}
\begin{align}
 \frac{\text{su}(N+1)_k \oplus \text{so}(2N)_1}{\text{su}(N)_{k+1} \oplus \text{u}(1)_{N(N+1)(N+k+1) }}
 \label{CPNmodel}
\end{align}
with the central charge 
\begin{align}
 c = \frac{3Nk}{k+N+1} \, .
\end{align}
Original proposal involves the 't Hooft limit,  where $N,k \to \infty$ with finite \eqref{thooft}.
It is claimed that the theory is a minimal model with respect to the ${\cal N}=(2,2)$ super $W_{N+1}$ algebra \cite{Ito:1990ac}.
The spectrum of the supergravity has been reproduced by the 't Hooft limit of the dual CFT \cite{Creutzig:2011fe,Candu:2012jq}. Boundary correlation functions are studied as well in \cite{Creutzig:2012xb,Moradi:2012xd}.
Recently, some classical geometry in the higher spin supergravity by $\text{sl}(N+1|N) \oplus \text{sl}(N+1|N)$ 
Chern-Simons theory has been investigated in \cite{Tan:2012xi,Datta:2012km,Peng:2012ae}.
In particular, conical defects in the Chern-Simons theory have been constructed in 
\cite{Tan:2012xi,Datta:2012km}. 
In this paper, we study the properties of the conical defects in more detail,
and we interpret them in terms of the dual CFT.

The rest of this paper is organized as follows; 
in the next section, we find out the conical defects in $\text{sl}(N+1|N) \oplus \text{sl}(N+1|N)$ 
Chern-Simons theory which preserve the maximal number of fermionic higher spin symmetry.
The conical geometry is classified by a $\text{SL}(N+1|N)$ holonomy matrix with
eigenvalues parametrized by integer numbers, and it can be
mapped to a smooth geometry by a gauge transformation. 
In section \ref{smooth}, we extend the class of smooth geometry with maximal supersymmetry
such that the interpretation in the dual CFT is possible. 
In section \ref{KSmodel} we identify the smooth solutions with 
primary states in the dual CFT.
We allow both anti-periodic and periodic boundary conditions for the Killing spinors along the 
spatial cycle of conical geometry, and each case corresponds to
NSNS-sector or RR-sector of the dual CFT. 
As a check we compare some charges of the $W$-algebra.
Furthermore, we map the null vectors in the dual CFT
to the residual higher spin symmetry
of the conical geometry by following the recent argument in \cite{Perlmutter:2012ds}.
Conclusion and discussions are given in section \ref{conclusion}.
In appendix \ref{superalgebra}, we summarize some useful formulas on $\text{sl}(N+1|N)$
Lie superalgebra.
In appendix \ref{degenerate} we examine degenerate representations of ${\cal N}=2$
$W_{N+1}$ algebra.

\subsection*{Note added}

While completing this work, the revised version of \cite{Datta:2012km} appeared in the arXiv.
The authors included study on fermionic symmetry of conical defects for higher spin supergravity
described by $\text{sl}(N+1|N) \oplus \text{sl}(N+1|N)$ Chern-Simons theory with $N \geq 3$,
while they dealt with only $N=2$ case in the previous version.
 There is overlap with the section \ref{CD} of this paper.

\section{Conical defects in higher spin supergravity}
\label{CD}

In \cite{Castro:2011iw} conical defects in a higher spin theory described
by $\text{sl}(N) \oplus  \text{sl}(N)$ Chern-Simons theory have been
studied and applied to the duality proposed in \cite{Gaberdiel:2010pz}.
The arguments are refined in \cite{Perlmutter:2012ds}.
In this section, we would like to investigate on the conical defect geometry in
a higher spin supergravity described by 
$\text{sl}(N+1|N)  \oplus  \text{sl}(N+1|N)$ Chern-Simons gauge theory.
We will apply the results to the ${\cal N}=2$ duality by \cite{Creutzig:2011fe}
in later sections.

\subsection{Conical defects}

We would like to consider $\text{sl}(N+1|N) \oplus  \text{sl}(N+1|N)$ Chern-Simons gauge 
theory. Its action is given by
\begin{align}
 S = S_\text{CS} [A] - S_\text{CS} [\tilde A] \, ,
 \label{CSaction}
\end{align}
where
\begin{align}
 S_\text{CS} [ A ] = \frac{\hat k}{4 \pi} \int \text{str}
\left( A \wedge dA + \frac{2}{3} A \wedge A \wedge A \right) \, .
\end{align}
Here the gauge fields take values in $\text{sl}(N+1|N)$ Lie superalgebra.%
\footnote{Some basics on the superlagebra including the definition of ``str'' may be found in appendix \ref{superalgebra}.     A review on superalgebras is given by \cite{Frappat:1996pb}.}
The action is invariant under the following gauge transformation as
\begin{align}
 \delta A = d \lambda + [A , \lambda] \, ,
 \qquad \delta \tilde A = d \tilde \lambda + [\tilde A , \tilde \lambda] \, .
 \label{gaugesym}
\end{align}
We define generalized dreibein and spin connection as
\begin{align}
 e = \frac{\ell}{2} (A - \tilde A) \, , \qquad
\omega =   \frac{1}{2} (A + \tilde A) \, .
\end{align}
We identify a $\text{sl}(2)$ subsector $\{L_0, L_{\pm 1}\}$ of $\text{sl}(N+1|N)$ as a 
gravitational sector. Then, the Chern-Simons level $\hat k$, AdS radius $\ell$ and
Newton's constant $G$ are related as
\begin{align}
 \hat k = \frac{\ell}{8G \epsilon_N} \, , \qquad
 \epsilon_N = \text{str} \, (L_{0} L_{0}) \, .
 \label{Newton}
\end{align}
With this notation the metric is 
\begin{align}
 g_{\mu \nu} = \frac{1}{\epsilon_N} \text{str} \, (e_\mu e_\nu) \, .
\end{align}
In this paper, we only use the superprincipal embedding of $\text{osp}(1|2)$ into $\text{sl}(N+1|N)$,
and in that case
\begin{align}
 \epsilon_N = \frac{N(N+1)}{4} \, .
\end{align}
See appendix \ref{superalgebra} (and also \cite{Frappat:1996pb}) for the details of the embedding.

We would like to study locally AdS$_3$ space with a conical singularity in a certain chosen gauge.
First we consider the ansatz for the gauge field configuration as \cite{Castro:2011iw,Tan:2012xi}
\begin{align}
 A = b^{-1} a_+ b dx^+ + L_0 d \rho \, , \qquad
  \tilde A = - b a_- b^{-1} dx^- - L_0 d \rho \, .
  \label{Aconfig}
\end{align}
The radial coordinate is $\rho$ and the light-like coordinates are $x^{\pm} = \phi \pm t $.
Here $b= \exp (\rho L_0) $ and 
\begin{align}
 a_+ = \sum_{k=1}^{N} B_k (a_k, b_k) + \sum_{\bar k =N+2}^{2N} B_{\bar k} (a_{\bar k} , b_{\bar k}) \, , \qquad
 a_- = - \sum_{k=1}^{N} B_k (c_k , d_k) - \sum_{\bar k =N+2}^{2N} B_{\bar k} (c_{\bar k} , d_{\bar k}) 
 \label{ansatz0}
\end{align}
with
\begin{align}
 [B_K (x,y) ]_{IJ} = x \delta_{I,K} \delta_{J,K+1} - y \delta_{I,K+1} \delta_{J,K} \, .
\end{align}
We use the Capital letters for $K,I,L=1,2,\ldots , 2N+1$, small letters for $k=1,2,\ldots,N+1$
and barred ones for $\bar k = N+2,N+2, \ldots , 2N+1$.
The arguments $a_L,b_L,c_L,d_L$ are  constant.
Assuming the static geometry with $g_{++} = g_{--}$ and the locally AdS metric, we have to set 
$a_L = b_L = c_L = d_L$ up to a similarity transformation.
The metric is now
\begin{align}
 \ell^{-2} ds^2 = d \rho ^2 - (e^\rho + M_N e^{-\rho})^2 dt^2 + 
 (e^\rho - M_N e^{- \rho})^2 d \phi^2  \label{conicaldf}
\end{align}
with
\begin{align}
 M_N =  \frac{1}{2 \epsilon_N } \left( \sum_{k=1}^{N} a_k^2 -   \sum_{\bar k=N+2}^{2N} a_{\bar k}^2 \right) \, .
 \label{M_N}
\end{align}
Here we redefine $\rho \to \rho + \ln \sqrt{ M_N }$.
The geometry has a conical singularity at $e^{2 \rho_0} = M_N$ with deficit angle $2\pi (1 - 2 \sqrt{M_N})$.

\subsection{Killing spinor equations}

We consider the gauge field configuration corresponding to a conical defect,
where in particular the fermionic components are set to be zero.
The fermionic higher spin symmetry is generated by the gauge 
transformation \eqref{gaugesym} which does not generate any non-zero fermionic components.
This condition is equivalent to the Killing spinor equation
\begin{align}
 {\cal D}_\mu \epsilon  \equiv \partial_\mu \epsilon + [A_\mu , \epsilon ] = 0 \, ,
 \label{Killing}
\end{align}
where $\epsilon$ is an odd element of $\text{sl}(N+1|N)$ superalgebra. 
Assuming that the supermatrices $a_+,a_-$ in \eqref{Aconfig} are diagonalizable, we can write
the ansatz \eqref{ansatz0} in the following form up to a bosonic gauge transformation
\begin{align}
 a_+ = \sum_{l=1}^{M} B_{2l-1} (a_{2l-1},a_{2l-1}) 
   + \sum_{\bar l = M+1}^{2M} B_{2\bar l} (a_{2 \bar l }, a_{2 \bar l }) 
   \label{ansatz1}
\end{align}
for $N=2M$ with $M \in \mathbb{Z}$ and 
\begin{align}
 a_+ = \sum_{l=1}^{M} B_{2l-1} (a_{2l-1},a_{2l-1}) 
   + \sum_{\bar l = M+1}^{2M-1} B_{2\bar l -1} (a_{2 \bar l  -1}, a_{2 \bar l - 1}) 
   \label{ansatz2}
\end{align}
for $N=2M-1$.
With the help of bosonic gauge transformation, we set $a_1 \geq a_3 \geq \cdots \geq a_{2M-1}$, and
$a_{2M+2} \geq a_{2M+4} \geq \cdots \geq a_{4M}$ for $N=2M$
and $a_{2M+1} \geq a_{2M+3} \geq \cdots \geq a_{4M-3}$ for $N=2M-1$.

The generators of $\text{sl}(N+1|N) $ Lie superalgebra can be represented in terms of
supermatrix, see appendix \ref{superalgebra}.
 Let us write basic $(2N+1) \times (2N+1)$ supermatrices as 
$(e_{IJ})_{KL} = \delta_{IK} \delta_{JL}$. The fermionic generators are then
given by $e_{i, \bar \jmath}$ and $e_{\bar \imath , j}$ where $i,j = 1, \ldots , N+1$ and
$\bar \imath , \bar \jmath = N+2 , \ldots , 2N+1$. Thus $\epsilon$ can be expanded as 
\begin{align}
 \epsilon = \sum_{i, \bar \jmath} \epsilon^{i, \bar \jmath} e_{i , \bar \jmath} + 
 \sum_{\bar \imath , j} \epsilon^{\bar \imath , j} e_{\bar \imath , j} \, .
 \label{killingspinor}
\end{align}
The bosonic gauge field configuration we are considering  does not mix  $e_{i, \bar \jmath}$ and $e_{\bar \imath , j}$, so we can safely set $\epsilon^{\bar \imath ,j} = 0$.
{}From the expression of $L_0$ in \eqref{L_n} and \eqref{K_0}, we have
\begin{align}
 [L_0 , e_{i , \bar \jmath} ] = ( - i + \bar \jmath - N - \tfrac12 ) e_{i, \bar \jmath} \, .
\end{align}
Thus the Killing spinor can be set as 
\begin{align}
 \epsilon_{i, \bar \jmath} = {\cal R} (\rho) \hat  \epsilon_{i, \bar \jmath} (x^+) \, , \qquad
  {\cal R} (\rho) = \exp ( (i - \bar \jmath + N + \tfrac12)  \rho ) \, .
\end{align}
For the $x^+$-dependence, we use the properties of generators as
\begin{align}
& [ B_{k} (a_{k}, a_{k}), e_{i , \bar \jmath}] = a_{k} (- \delta_{i,k} e_{i+1, \bar \jmath} + \delta_{i,k+1} e_{i-1 , \bar \jmath}) \, , \\
& [ B_{\bar k} (a_{\bar k}, a_{\bar k}), e_{i , \bar \jmath}] = a_{\bar k} (- \delta_{\bar \jmath ,\bar k  } e_{i , \bar \jmath + 1} + \delta_{\bar \jmath,\bar k+1} e_{i, \bar \jmath - 1} ) \, .
\end{align}
For $N=2 M$, this yields
\begin{align}
 & [ B_{2l-1} (a_{2l-1}, a_{2l-1}), e_{2p-1 , \bar \jmath} \pm i e_{2p , \bar \jmath} ] = \pm i a_{2l-1} \delta_{l,p}(e_{2p-1 , \bar \jmath} \pm i e_{2p , \bar \jmath}) \, , \\
& [ B_{2 \bar l} (a_{2 \bar l}, a_{2 \bar l}), e_{i , 2 \bar p} \pm i  e_{i , 2 \bar p + 1} ] = \pm i a_{2\bar l} \delta_{\bar l,\bar p}(e_{i, 2 \bar p } \pm i e_{i, 2 \bar p +1}) 
\end{align}
for $l,p=1,2,\ldots, M$ and $\bar l , \bar p = M+1 , M+2 , \ldots , 2M$.
The eigenvectors can be constructed as
\begin{align}
 E^{\eta_p , \bar \eta_{\bar l}}_{p , \bar l} = 
 e_{2p-1 , 2 \bar l} + i \eta_{p} e_{2p , 2 \bar l }
 + i \bar \eta_{\bar l} ( e_{2p-1 , 2 \bar l + 1} + i \eta_{p} e_{2p , 2 \bar l +1} )
\end{align}
with $\eta_p , \bar \eta_{\bar l} = \pm 1$, whose eigenvalue is
\begin{align}
 [a_+ ,  E^{\eta_p , \bar \eta_{\bar l}}_{p , \bar l}]
  = i ( \eta_{p} a_{2p - 1} + \bar \eta_{\bar l} a_{2 \bar l} )   E^{\eta_p , \bar \eta_{\bar l}}_{p , \bar l} \, .
\end{align}
We have another set of eigenvectors as
\begin{align}
 E^{\bar \eta_{\bar L}}_{ \bar l} = e_{N+1, 2\bar l} + i \bar \eta_{\bar l} e_{N+1, 2\bar l + 1} \, , \qquad
 [ a_+ , E^{\bar \eta_{\bar L}}_{ \bar l} ] = i \bar \eta_{\bar l} a_{2 \bar l} E^{\bar \eta_{\bar L}}_{ \bar l} 
\end{align}
for $\bar l  = M+1 , M+2 , \ldots , 2M$.
Therefore, the solutions to the Killing spinor equation are  given by
\begin{align}
 {\cal R} (\rho) ^{-1} \epsilon 
 =  \sum_{ p =1}^M \sum_{ \bar l = M+1}^{2M}
  \sum_{ \eta_p , \bar \eta_{\bar l} = \pm 1}  c^{\eta_p , \bar \eta_{\bar l}}_{p , \bar l}
  e^{-i ( \eta_{p} a_{2p - 1} + \bar \eta_{\bar l} a_{2 \bar l} ) x^+ }E^{\eta_p , \bar \eta_{\bar l}}_{p , \bar l}  \\
 + \sum_{ \bar l = M+1}^{2M}\sum_{ \bar \eta_{\bar l} = \pm 1}
  c^{ \bar \eta_{\bar l}}_{\bar l}
  e^{-i  \bar \eta_{\bar l} a_{2 \bar l}  x^+ }E^{\bar \eta_{\bar l}}_{\bar l}  \nonumber
\end{align}
with constants $c^{\eta_p , \bar \eta_{\bar l}}_{p , \bar l}, c^{ \bar \eta_{\bar l}}_{\bar l}$.
 In the same way, we have for $N$ odd
\begin{align}
 {\cal R} (\rho) ^{-1} \epsilon 
 =  \sum_{ p =1}^M \sum_{ \bar l = M+1}^{2M-1}
  \sum_{ \eta_p , \bar \eta_{\bar l} = \pm 1}  c^{\eta_p , \bar \eta_{\bar l}}_{p , \bar l}
  e^{-i ( \eta_{p} a_{2p - 1} + \bar \eta_{\bar l} a_{2 \bar l-1} ) x^+ }E^{\eta_p , \bar \eta_{\bar l}}_{p , \bar l}  \\
+ \sum_{ p=1}^{M}\sum_{ \eta_p = \pm 1}
  c^{ \eta_{p}}_{p}
  e^{-i  \eta_{p} a_{2 p - 1}  x^+ }E^{\eta_{p}}_{p}  \, , \nonumber
\end{align}
 where
 \begin{align}
 &E^{\eta_p , \bar \eta_{\bar l}}_{p , \bar l} = 
 e_{2p-1 , 2 \bar l-1} + i \eta_{p} e_{2p , 2 \bar l -1}
 + i \bar \eta_{\bar l} ( e_{2p-1 , 2 \bar l } + i \eta_{p} e_{2p , 2 \bar l } ) \, , \\
  &E^{\eta_p }_{p } =  e_{2p-1 , 2N+1} + i \eta_{p} e_{2p , 2N + 1} 
\end{align}
with constants $c^{\eta_p , \bar \eta_{\bar l}}_{p , \bar l}, c^{ \eta_{p}}_{p}$.

If a part of supersymmetry is preserved, then the corresponding Killing spinors have to satisfy 
anti-periodic boundary condition around the $\phi$-cycle. Here we would like to require the maximal number of supersymmetry, thus we should have $N(N+1)$ Killing spinors $\epsilon^{i, \bar \jmath}$ for all $i, \bar \jmath$.
This leads to the condition that $a_{2l-1} = p_l$ with $p_l  \in \mathbb{Z}$ and 
$a_{2 \bar \jmath} = q_{\bar \jmath} + 1/2$ with $q_{\bar \jmath} \in \mathbb{Z} $ for $N=2M$, and
$a_{2l-1} = p_l + 1/2$ with $p_l  \in \mathbb{Z} $ and 
$a_{2 \bar \jmath - 1} = q_{\bar \jmath} $ with $q_{\bar \jmath} \in \mathbb{Z} $ for $N=2M-1$. 
Notice that this condition coincides with the requirement that the holonomy matrix along the 
$\phi$-cycle is 
\begin{align}
 \text{Hol}_\phi (A)=  \exp \left(\oint A_\phi d\phi \right) 
  =  (-1)^N \mathbf{1}_{\text{sl} (N+1)} 
  \otimes (-1)^{N-1} \mathbf{1}_{\text{sl}(N)}
  \otimes  \mathbf{1}_{\text{u}(1)} 
\end{align}
up to a similarity transformation. It is a center of the bosonic subalgebra 
$\text{sl}(N+1) \oplus \text{sl}(N) \oplus \text{u}(1)$.
In general, the notion of singularity is not gauge invariant in higher spin gauge theory.
Since the holonomy matrix is an gauge invariant operator, the trivial
holonomy matrix suggests that our geometry is actually singularity free.
In fact, by applying a gauge transformation as in (3.31) of \cite{Castro:2011iw},
we can map the conical defect geometry in \eqref{conicaldf} to a smooth wormhole
geometry.

We can also think of geometry with Killing spinors satisfying periodic boundary condition
around the $\phi$-cycle.  Assuming the maximal number of Killing spinors,
we have to set that  $a_{2l-1} = p_l$ with $p_l  \in \mathbb{Z}$ and 
$a_{2 \bar \jmath} = q_{\bar \jmath} $ for $N=2M$
or $a_{2 \bar \jmath-1} = q_{\bar \jmath} $ for $N=2M-1$ with $q_{\bar \jmath} \in \mathbb{Z}$.
The holonomy matrix along the $\phi$-cycle is  
\begin{align}
 \text{Hol}_\phi (A) =  \mathbf{1}_{\text{sl} (N+1)} 
  \otimes \mathbf{1}_{\text{sl}(N)}
  \otimes  \mathbf{1}_{\text{u}(1)} 
\end{align}
up to a similarity transformation. This implies that the geometry is singularity free,
and again we can map the conical defect geometry into a smooth wormhole geometry
by a gauge transformation. 

The AdS space corresponds to 
\begin{align}
 a_{2l - 1} = M + 1 - l \, , \qquad a_{2 \bar \jmath} = 2 M - \bar \jmath + \tfrac12
\end{align}
 for $N=2M$ and 
\begin{align}
 a_{2l - 1} = M + \tfrac12 - l \, , \qquad a_{2 \bar \jmath - 1} = 2 M - \bar \jmath 
\end{align}
for $N=2M-1$. Both lead to  $M_N = 1/4$ in \eqref{M_N}.
For geometry with conical defect, we have a condition as
 \begin{align}
  0 < M_N <  \tfrac14 ~.
 \end{align}
Let us check whether there are configurations satisfying this condition for small $N$ cases,
see also \cite{Tan:2012xi,Datta:2012km}.
For $N=1,2$, we can see that there is no such a solution.
For $N=3$, non-trivial solutions are 
\begin{align}
(a_1,a_3,a_5)=(\tfrac32,\tfrac32,2),(\tfrac12,\tfrac12,0), (\tfrac12,0,0) \, \qquad M_3 = \tfrac{1}{12},
\tfrac{1}{12},\tfrac{1}{24}
\end{align}
for anti-periodic case and 
\begin{align}
(a_1,a_3,a_5)=(p_1,1,p_1),(1,0,0) \, \qquad M_3 = \tfrac{1}{6},\tfrac{1}{6}
\end{align}
with $p_1 \in \mathbb{Z}$ for periodic case. Possibly there are more.
For larger $N$, we can easily find out more solutions.

\section{Smooth geometry with maximal supersymmetry}
\label{smooth}

In the previous section, we have dealt with conical defects with $0 < M_N < 1/4$ for \eqref{M_N},
and the solutions may be not physical outside the region.
However,  we will later compare classical gravity solutions 
to primary states in a non-unitary CFT. Thus we do not need to focus on physical 
solutions and  remove the restriction from now on.
Furthermore, we move to the Euclidean model given by the gauge fields $A, \tilde A$ with
taking complex values and $A^\dagger = - \tilde A$. 
In this section, we examine with this setup more generic solutions which are not included in the ansatz \eqref{ansatz1} or \eqref{ansatz2}.
In the next section, we will see that  the map from these solutions  
to CFT primary states works  very nicely as in the bosonic case \cite{Castro:2011iw,Perlmutter:2012ds}.

\subsection{Killing spinors and holonomy matrix}
 
Performing the Wick rotation to \eqref{Aconfig},
we consider the gauge field configuration with
\begin{align}
 A = b^{-1} a_+ b d w + L_0 d \rho \, , \qquad
  \tilde A = - b a_- b^{-1} d \bar w  - L_0 d \rho \, ,
  \label{Aconfigz}
\end{align}
where $w = \phi + i \tau$ and $\bar w = \phi - i \tau$.
Here we assume that $a_+$ and $a_-$ are elements of $\text{sl}(N+1|N)$ superalgebra with
complex values and they can be diagonalized by some supermatrix.
Moreover, we set $A^\dagger = - \tilde A$ and $b= \exp (\rho L_0) $.

As argued in \cite{Datta:2012km}, solutions to the killing spinor equation \eqref{Killing}
may be written as
\begin{align}
 \epsilon (x) = {\cal P} \exp (- \int_{x_0}^x A_\mu d x^\mu) \hat \epsilon (x_0) 
 {\cal P} \exp (\int_{x_0}^x A_\mu d x^\mu) \, .
\end{align}
The problem is to find out the gauge field configuration such that the maximal number of Killing spinors
satisfy  anti-periodic or periodic boundary condition around the $\phi$-cycle.
When going around the cycle, the factor becomes holonomy matrix as
\begin{align}
 \text{Hol}_\phi (A) =  \exp \left(\oint A_\phi d\phi \right) 
  = S^{-1} \exp \left(2 \pi \sum_{I=1}^{2N+1} \theta_I e_{II} \right) S  \, ,
  \label{eigenvalue}
\end{align}
where $S$ is a supermatrix depending on $\rho$.
Change the basis of spinor as 
 \begin{align}
  \epsilon (x_0) = S \hat \epsilon (x_0)  S^{-1}
  = \sum_{l = 1}^{N+1} \sum_{\bar \jmath = N+2}^{2N+1} \epsilon^{l, \bar \jmath} e_{l, \bar \jmath} \, ,
 \end{align}
we can see that when the spinor with only $\epsilon^{l, \bar \jmath} \neq 0$ goes around the
cycle the phase factor becomes  
 \begin{align}
  \exp \left( - 2 \pi \sum_{I=1}^{2N+1} \theta_I e_{II} \right) e_{l , \bar \jmath }  \exp \left( 2 \pi \sum_{I=1}^{2N+1}  \theta_I e_{II} \right)
   = \exp \left( - 2 \pi (  \theta_l - \theta_{\bar \jmath}) \right)  e_{l, \bar \jmath} \, .
   \label{spinorphase}
 \end{align}
The phase factor should be $-1$ for all possible set of $(l,\bar \jmath)$ when all Killing spinors satisfy the
anti-periodic boundary condition. In the same way, the phase factor should be $+1$ for all 
$(l,\bar \jmath)$ when all spinors satisfy the periodic boundary condition.

The condition of anti-periodicity for all Killing spinors is thus
 \begin{align}
  \theta_l - \theta_{\bar \jmath} \in i (\mathbb{Z} + \tfrac12 ) 
 \end{align}
 for all $l, \bar \jmath$.
Generic solutions are
\begin{align}
 \theta_l = i ( p_l + \beta )\, , \qquad \theta_{\bar \jmath} = i (q_{\bar \jmath} + \tfrac12 + \beta )
\end{align}
 with $p_l , q_{\bar \jmath} \in \mathbb{Z}$. However,  the supertraceless condition of
 $\text{sl}(N+1|N)$ reads
 \begin{align}
  \beta = - \sum_{l=1}^{N+1} p_l + \sum_{\bar \jmath = N+2 }^{2N+1}( q_{\bar \jmath} + \tfrac12) \, ,
  \label{beta}
 \end{align} 
 which is integer for $N=2M$ and half-integer for $N=2M-1$. 
 Thus we see
  \begin{align}
  \theta_l = i \left(p'_l + \frac{1-(-1)^N}{4} \right) \, , \qquad
  \theta_{\bar \jmath} = i \left ( q'_{\bar \jmath} + \frac{1 + (-1)^N}{4} \right) \, ,
 \end{align}
where $q'_l, p'_j \in \mathbb{Z}$. 
 It is convenient to write down $\theta_L$ in terms of bosonic subalgebra
 $\text{sl}(N+1) \oplus \text{sl} (N)  \oplus \text{u}(1)$ as
\begin{align}
 &- i \theta_i
       = \tilde l^{(1)}_i + \rho^{(1)}_i + \frac{\tilde m}{N+1} 
       = \tilde r^{(1)}_i - \frac{|\tilde \Lambda^{(1)}|}{N+1} + \frac{N+2}{2} - i + \frac{\tilde m}{N+1} 
\, , \label{Young1} \\
 &- i \theta_{N+1+j} 
     = \tilde l^{(2)}_j + \rho^{(2)}_j + \frac{\tilde m}{N} 
     = \tilde r^{(2)}_j - \frac{|\tilde \Lambda^{(2)}|}{N} + \frac{N+1}{2} - j + \frac{\tilde m}{N} \, .
     \label{Young2}
\end{align}
Here $\sum_i \tilde l^{(a)}_i = 0$,  $\tilde r_i ^{(a)} \in \mathbb{Z}$ and 
$|\tilde \Lambda^{(a)}| = \sum_i \tilde r_i^{(a)}$.
The Weyl vectors $\rho^{(a)}_i$ are defined in \eqref{rho1} and \eqref{rho2}.
{}From the condition for $\beta$, we find that 
\begin{align} \label{codtm1}
  \tilde m \in 
 - N |\tilde \Lambda^{(1)}| + (N+1) |\tilde \Lambda^{(2)}| + N (N+1) \mathbb{Z} \, .
 \end{align}
Holonomy matrix is now
 \begin{align}
 \text{Hol}_\phi (A) = e^{2\pi i (\frac{N}{2} + \frac{\tilde m}{N+1})} \mathbf{1}_{\text{sl} (N+1)} 
  \otimes e^{2\pi i (\frac{N+1}{2} - \frac{\tilde m}{N})}\mathbf{1}_{\text{sl}(N)}
  \otimes e^{ - 2\pi i \frac{\tilde m }{ N(N+1)}} \mathbf{1}_{\text{u}(1)} 
  \label{holNS}
\end{align}
up to a similarity transformation. Notice that it is a center of bosonic subalgebra 
$\text{sl}(N+1)  \oplus \text{sl} (N)  \oplus \text{u}(1)$ with complex elements,
thus the configurations considered should correspond to smooth geometries in some gauge choice.

Similarly, for the periodic case we need to assign
 \begin{align}
  \theta_l- \theta_{\bar \jmath} \in i \mathbb{Z} 
   \end{align}
for all $l, \bar \jmath$, and solutions are
\begin{align}
 \theta_l = i (p_l + \beta )\, , \qquad \theta_{\bar \jmath} = i (q_{\bar \jmath} + \beta )
\end{align}
 with $p_l , q_{\bar \jmath} \in \mathbb{Z}$. The supertraceless condition leads
 \begin{align}
  \beta = - \sum_{l=1}^{N+1} p_l + \sum_{\bar \jmath = N+2 }^{2N+1} q_{\bar \jmath}  \, ,
 \end{align} 
 which is also an integer number. 
Thus we may define
\begin{align} 
 &- i \theta_i
       = \tilde l^{(1)}_i + \rho^{(1)}_i + \frac{\tilde m}{N+1} 
       = \tilde r^{(1)}_i - \frac{|\tilde \Lambda^{(1)}|}{N+1} + \frac{N+2}{2} - i + \frac{\tilde m}{N+1} 
\, , \label{Young3} \\
 & - i \theta_{N+1+j}
     = \tilde l^{(2)}_j + \rho^{(2)}_j + \frac{\tilde m}{N} 
     = \tilde r^{(2)}_j - \frac{|\tilde \Lambda^{(2)}|}{N} + \frac{N+1}{2} - j + \frac{\tilde m}{N} \, , \label{Young4}
\end{align}
with $\sum_i \tilde l^{(a)}_i = 0$ and $\tilde r_i ^{(a)} \in \mathbb{Z}$.
{}From the condition for $\beta$, we have 
\begin{align} \label{codtm2}
  \tilde m \in 
 - N |\tilde \Lambda^{(1)}| + (N+1) |\tilde \Lambda^{(2)}| + N (N+1) ( \mathbb{Z} + \tfrac12 ) \, . 
 \end{align}
The holonomy matrix is
 \begin{align}
 \text{Hol}_\phi (A) = e^{2\pi i \frac{\tilde m}{N+1}} \mathbf{1}_{\text{sl} (N+1)} 
  \otimes e^{- 2\pi i \frac{\tilde m}{N}}\mathbf{1}_{\text{sl}(N)}
  \otimes e^{- 2\pi i \frac{\tilde m }{ N(N+1)}} \mathbf{1}_{\text{u}(1)} 
  \label{holR}
\end{align}
up to a similarity transformation. 

\subsection{Asymptotically AdS geometry}

In order to compare the supergravity with the dual CFT, we have to search solutions which 
approach to AdS space at $\rho \to \infty$. For higher spin gauge theory, we need to assign
boundary condition also for higher spin fields, which can be expressed as 
\cite{Campoleoni:2010zq}
\begin{align}
  ( A - A_\text{AdS} ) |_{\rho \to \infty} \sim {\cal O} (1) \, .
  \label{DS}
\end{align}
The gauge field configuration $A_\text{AdS}$ corresponding to the AdS background is 
given by \eqref{Aconfigz} with $a_+ =  L_{1}$ and $a_- = - L_{-1}$.
The condition is shown to be equivalent to the Drinfeld-Sokolov reduction in \cite{Campoleoni:2010zq},
and in our case, the classical asymptotic symmetry under the condition is ${\cal N}=(2,2)$ super 
$W_{N+1}$ algebra \cite{Creutzig:2011fe}.

The conical geometry considered has higher spin charges associated with the $W$-algebra, 
and we would like to compute them in this subsection. In order to assign the  asymptotic
boundary condition to the gauge fields, it is convenient to decompose the $\text{sl}(N+1|N)$
elements by its $\text{sl}(2)$ subalgebra. The decomposition depends on how we embed 
$\text{sl}(2)$, and we have chosen the one in \cite{Creutzig:2011fe} such that
\begin{align}
  \text{sl}(N+1|N)= \text{sl}(2) \oplus ({\oplus}_{s=3}^{N+1} g^{(s)}) \oplus
   ({\oplus}_{s=1}^N g^{(s)}) \oplus 2 \cdot ({\oplus}_{s=1}^N g^{(s+1/2)} ) \, ,
\end{align}
where $g^{(s)}$ is the $(2s-1)$-dimensional representation of $\text{sl}(2)$.
Notice that the integer spin elements are even and the half-integer spin elements are odd
with respect to the $\mathbb{Z}_2$-grading of the superalgebra.
With this decomposition, the generators may be given by
\begin{align}
 V^{(s)+}_n ~ (s=2,3,\ldots, N+1) \, , \quad V^{(s)-}_n ~ (s=1,2,\ldots , N) \, , \quad F^{(s)\pm}_r ~
 (s=1,2,\ldots , N) \, ,
\end{align}
where $|n| \leq s-1, |r| \leq s-1/2$. The embedded sl(2) is generated by 
$L_m = V^{(2)+}_m$ with $m=0,\pm1$.
Some of the commutation relations may be found in appendix 
\ref{superalgebra}.

In terms of these generators, the gauge field configuration satisfying the asymptotic AdS condition
\eqref{DS} can be set as
\begin{align}
 a_+ (t + \theta) = L_1 & + \frac{1}{\hat k} \Biggl ( \sum_{s \geq 2} \frac{1}{N_s^+}L^{+}_s (t + \theta) V^{(s)+}_{-s+1} + 
      \sum_{s \geq 1} \frac{1}{N_s^-} L_s^- (t + \theta ) V^{(s)-}_{-s+1}  \label{Wgenerators_g} \\
       & + \sum_{s \geq 1} \frac{1}{M_{s+1/2}^+} G^+_{s+1/2} (t + \theta) F^{(s)+}_{-s + 1/2} 
       + \sum_{s \geq 1 } \frac{1}{M_{s+1/2}^+}G^-_{s+1/2} (t + \theta) F^{(s)-}_{-s + 1/2} \Biggr ) \nonumber
\end{align}
by utilizing residual gauge transformation. 
Here we have defined
\begin{align}
 N_s^\pm = \text{str}\, (V^{(s)\pm}_{s-1} V^{(s)\pm}_{-s+1}) \, , \qquad
 M_{s+1/2}^\pm = \text{str} \, (F^{(s)\pm}_{s-1/2} F^{(s)\pm}_{-s+1/2}) \, .
\end{align}
At the boundary, the functions $L^{\pm}_s (\theta), G^\pm_{s+1/2}(\theta)$ 
act as generators of classical ${\cal N}=2$ super $W_{N+1}$ algebra, see
 \cite{Creutzig:2011fe,Henneaux:2012ny,Hanaki:2012yf}.
In particular, the energy momentum tensor comes from $L^+_2 (\theta)$ and
the central charge is 
\begin{align}
 c = 12 \hat k \epsilon_ N = \frac{3  \ell}{2G} 
 \label{center_g}
\end{align}
in terms of parameters in \eqref{Newton}.
Note that this value is the same as the one obtained for
the pure gravity in \cite{Brown:1986nw}.
The other generators are primary with respect to the energy momentum tensor.

As for our geometry, we assume the form of the gauge field as \eqref{Aconfigz}, 
where $a_+$ takes a value in constant $\text{sl}(N+1|N)$ superalgebra.
It is useful to define
\begin{align}
  T^{(s) \pm}_n = \frac{1}{2} (V^{(s)+}_n \pm V^{(s)-}_n) \, , \qquad
   T^{(1)-}_0 = V^{(1)-}_0 \, , \qquad T^{(N+1)+}_n = V^{(N+1)+}_n \, ,
\end{align}
where $s=2,3,\ldots, N$. In this notation, 
$ T^{(s)+}_n$, $T^{(s)-}_n$ $(s \geq 2)$ and $ T^{(1)-}_0 $ generate $\text{sl}(N+1)$, $\text{sl}(N)$ and $\text{u}(1)$
bosonic subalgebras. Assigning the asymptotic AdS condition \eqref{DS}, the gauge field
takes the form of
\begin{align}
 a_+ (t + \theta) = L_1 & + \sum_{s \geq 2} \frac{ \hat k^{-s/2} }{t^{(s)}_+} v^{(s)}_+  T^{(s)+}_{-s+1} + 
    \sum_{s \geq 1} \frac{  \hat k^{-s/2} }{t^{(s)}_-} v^{(s)}_-  T^{(s)-}_{-s+1}
   \label{DSw}
\end{align}
with
\begin{align}
 t^{(s)}_\pm = \text{str}\,  (T^{(s)\pm}_{s-1} T^{(s)\pm}_{-s+1}) \, .
\end{align}
The constant coefficients $v^{(s)}_\pm$ are related to eigenvalues $\theta_L$ in \eqref{eigenvalue}
by a gauge transformation, and they correspond to the charges of ${\cal N}=2$ super $W_{N+1}$ algebra.
Notice that the fermionic components are set to be zero in our gauge configurations.
The above form with the normalization is particularly useful since we can just apply the result of 
\cite{Castro:2011iw} to the bosonic subalgebras.
The $\text{u}(1)$ charge can be easily read off as
\begin{align}
 v^{(1)}_- = -  i \hat k^{1/2} \tilde m
 \label{chargeu1}
\end{align}
by using the notation in appendix \ref{superalgebra}.
The other first few charges are \cite{Castro:2011iw} 
\begin{align}
 & v^{(2)}_+ = - \hat k C^{+}_2 (\tilde n^{(1)})  \, , \nonumber \\
 & v^{(3)}_+ = - i \hat k^{3/2} C_3^{+} (\tilde n^{(1)} ) \, , \label{charget+} \\
 & v^{(4)}_+ = \hat k^2 \left(  C_4^{+} (\tilde n^{(1)} ) - \frac{C^+_4 (\rho^{(1)})}{(C_2^+ (\rho^{(1)}))^2}
  ( C_2^+ (\tilde n^{(1)} ) ) ^2 \right) \, , \nonumber 
\end{align}
and 
\begin{align}
 & v^{(2)}_- = \hat k C^{-}_2 (\tilde n^{(2)})  \, , \nonumber \\
 & v^{(3)}_- =  i \hat k^{3/2} C_3^{-} (\tilde n^{(2)} ) \, , \label{charget-} \\
 & v^{(4)}_- = - \hat k^2 \left(  C_4^{-} (\tilde n^{(2)} ) - \frac{C^-_4 (\rho^{(2)})}{(C_2^- (\rho^{(2)}))^2}
  ( C_2^- (\tilde n^{(2)}) ) ^2 \right) \, , \nonumber 
\end{align}
where
\begin{align}
 C^+_s (\tilde n^{(1)}) = \frac{1}{s} \sum_{j=1}^{N+1} (\tilde n^{(1)}_j)^s \, , \qquad
 C^-_s (\tilde n^{(2)})= \frac{1}{s} \sum_{j=1}^{N} (\tilde n^{(2)}_j )^s \, 
\end{align}
with
\begin{align}
 \tilde n^{(a)}_j = \tilde l^{(a)}_j + \rho^{(a)}_j \, .
 \label{ntilde}
\end{align}
Notice that this expression holds both for anti-periodic and periodic cases.

In the previous subsection, we assumed that the supermatrix $a_+$ is diagonalizable. 
As discussed in \cite{Castro:2011iw}, the supermatrix of the form \eqref{DSw} can de 
diagonalized by supermatrix corresponding to $\text{sl}(N+1) \oplus \text{sl}(N)$ bosonic
subalgebra only when all $\tilde n^{(a)}_j$ are distinct for both $a=1,2$.
Thus we need to require
\begin{align}
 \tilde n^{(1)}_1 > \tilde n^{(1)}_2 > \cdots > \tilde n^{(1)}_{N+1} \, , \qquad
 \tilde n^{(2)}_1 > \tilde n^{(2)}_2 > \cdots > \tilde n^{(2)}_{N} \, .
\end{align}
In this case the holonomy matrix can be labeled by two Young diagrams $\Lambda^{(a)}$
along with the $\text{u}(1)$ charge $\tilde m$. In terms of parameters in 
\eqref{Young1}, \eqref{Young2}, \eqref{Young3} and \eqref{Young4},
the Young diagram $\Lambda^{(a)}$ has $\tilde r^{(a)}_j$ boxes in the $j$-th row.

\section{Relation to the $\mathbb{CP}^N$ model}
\label{KSmodel}

In \cite{Creutzig:2011fe} it was proposed that the ${\cal N}=2$ higher spin supergravity in \cite{Prokushkin:1998bq} is dual to the 't Hooft limit \eqref{thooft} of the $\mathbb{CP}^N$ Kazama-Suzuki model \eqref{CPNmodel}
\begin{align}
  \frac{\text{su}(N+1)_k \oplus \text{so}(2N)_1 }{\text{su}(N)_{k+1} \oplus \text{u}(1)_{N(N+1)(k+N+1)}} 
  \label{CPNmodel2}
\end{align}
with the central charge
\begin{align}
 c = \frac{3Nk}{k+N+1} \, .
\end{align}
The massless sector of the supergravity is described by 
$\text{shs}[\lambda] \oplus \text{shs}[\lambda]$ Chern-Simons theory, and the 
higher spin superalgebra $\text{shs}[\lambda]$ can be truncated to $\text{sl}(N+1|N)$
at $\lambda = N + 1 $.  
{}Based on the duality, we identify the classical smooth geometry of
 $\text{sl}(N+1|N) \oplus \text{sl} (N+1|N)$ Chern-Simons theory  considered in the 
previous section as a primary state in its dual CFT.
We perform several checks of this identification.
For the bosonic case, see \cite{Castro:2011iw,Perlmutter:2012ds}.

\subsection{Relation to primary states}

In order to compare with the $\text{sl}(N+1|N) \oplus \text{sl} (N+1|N)$ Chern-Simons theory,
we will find that it is necessary to shift the central charge $c$ of the Kazama-Suzuki model \eqref{CPNmodel2}
into not a physically allowed region. 
This implies that we need to move to a more 
generic theory with the same symmetry.
However, in this subsection, we still study the Kazama-Suzuki model since 
the difference appears from the next leading order of $1/c$ with large $c$ as we will see below.

The states of the Kazama-Suzuki model 
are labeled by $(\Lambda^{(1)} , \omega ; \Lambda^{(2)},m)$.
Here $\Lambda^{(1)}, \Lambda^{(2)}$ are highest weights of $\text{su}(N+1),\text{su}(N)$ and 
the $\text{u}(1)$ charge takes a value in $m \in \mathbb{Z}_\kappa $ with $\kappa = N(N+1)(k+N+1)$.
There are four representations of affine $\text{so}(2N)_1$ with $\omega = - 1 , 0 , 1, 2$.
Here $\omega =0 $ and $\omega = 2$ correspond to identity and vector representations,
respectively, and the fermions are in the NS-sector with anti-periodic boundary condition.
On the other hand, $\omega = -1$ and $\omega = 1$ correspond to co-spinor and spinor
representations, and the fermions are in the R-sector with periodic boundary condition.
The states of the coset are then obtained by the decomposition
\begin{align}
 \Lambda^{(1)} \otimes \omega  = 
  \oplus_{\Lambda^{(2)},m} (\Lambda^{(1)}, \omega ; \Lambda^{(2)},m) \otimes
   \Lambda^{(2)} \otimes m \, .
\end{align}
{}From the condition that the decomposition is possible, we have  a selection rule 
\begin{align} \label{selection}
 \frac{|\Lambda^{(1)}|}{N+1} - \frac{|\Lambda^{(2)}|}{N} + \frac{m}{N(N+1)} + \frac{\omega}{2} 
 = 0 ~ \text{mod} ~ 1 \, ,
\end{align}
where $|\Lambda^{(a)}|$ is the number of boxes of Young diagram corresponding to $\Lambda^{(a)}$.
See appendix \ref{superalgebra} for the notations.
In general, we should take case of field identification \cite{Gepner:1989jq} as well, but it is not 
relevant for our purpose.%
\footnote{
We should take the following identification among the states as 
$(\Lambda^{(1)},\omega ; \Lambda^{(2)},m) \simeq 
(A_{N+1} \Lambda^{(1)},\omega+2; A_N \Lambda^{(2)},m+k+N+1)$, where $A_M$ 
is an outer automorphism of $\text{su}(M)$. Later we consider an analytic continuation
on $k$, and the field identification does not make sense with an irrational $k$.}

The conformal weight of the primary state is in the NS-sector
\begin{align}
 h (\Lambda^{(1)}, \omega ; \Lambda^{(2)} , m) = n + \frac{\omega}{4} + \frac{1}{k + N + 1}
 \left (C^{(1)} (\Lambda^{(1)}) - C^{(2)} (\Lambda^{(2)}) - \frac{m^2}{2N(N+1)} \right)  \, ,
 \label{confcft0}
\end{align}
where $C^{(a)} (\Lambda^{(a)})$ is the second Casimir 
in the representation $\Lambda^{(a)}$ of $\text{su}(N+1)$ for $a=1$ and  $\text{su}(N)$ for $a=2$.
Here $n$ is the grade at which $(\Lambda^{(2)},m)$ appears as a descendant of $(\Lambda^{(1)}, \omega)$ (see, e.g., \cite{CFT}).
 The $\text{u}(1)$ charge is
\begin{align}
 q (\Lambda^{(1)}, \omega ; \Lambda^{(2)} , m) = 2 n'
 + \frac{\omega}{2} - \frac{m}{N+k+1} 
 \label{qcft0}
\end{align}
with an integer $n'$.
In the R-sector, we have
\begin{align}
 &h (\Lambda^{(1)}, \omega ; \Lambda^{(2)} , m) = n + \frac{N}{8} + \frac{1}{k + N + 1}
 \left (C^{(1)} (\Lambda^{(1)}) - C^{(2)} (\Lambda^{(2)}) - \frac{m^2}{2N(N+1)} \right )  \, , \nonumber \\
 &q (\Lambda^{(1)}, \omega ; \Lambda^{(2)} , m) = 2 n'
 + \frac{N}{2} + \frac{\omega - 1}{2} - \frac{m}{N+k+1} 
\end{align}
with some integers $n,n'$.

We would like to compare these primary states to the classical solutions of 
$\text{sl}(N+1|N) \oplus \text{sl}(N+1|N)$ Chern-Simons theory, where the 
classical limit corresponds to the limit with large Chern-Simons level $\hat k$.
It is also the same as the limit with large central charge $c$ for the asymptotic symmetry algebra as in \eqref{center_g}.
Thus, we should take the large central charge limit with $c \to \infty$ but with
$N$ kept finite. This implies that we need to consider an analytic continuation 
of $k$ to an unphysical value as
\begin{align}
 k = - (N+1) - \frac{3N(N+1)}{c} + { \cal O } (c^{-2}) \, .
\end{align}
The validity of the analytic continuation is discussed in \cite{Candu:2012tr},
see also \cite{Gaberdiel:2012ku,Candu:2012ne} for bosonic cases.
With this limit, the conformal weight and the $\text{u}(1)$ charge become
\begin{align}
 &h (\Lambda^{(1)}, \omega ; \Lambda^{(2)} , m) = - \frac{c}{12 \epsilon_N}
 \left( C^{(1)} (\Lambda^{(1)}) - C^{(2)} (\Lambda^{(2)}) - \frac{m^2}{2N(N+1)} \right) \, 
 \label{confcft} ,\\
 &q (\Lambda^{(1)}, \omega ; \Lambda^{(2)} , m) =  \frac{c}{12 \epsilon_N} m 
\nonumber
\end{align}
for all choices of $\omega$.

In the previous section, we found a set of smooth geometry preserving the maximal
number of fermionic symmetry. It is classified by the holonomy matrix with eigenvalues
$ \theta_L$ $(L=1,2,\ldots, 2N+1)$, which are parametrized by two Young diagrams
$\tilde \Lambda^{(a)}$ $(a=1,2)$ and one integer $\tilde m$ as in 
\eqref{Young1}, \eqref{Young2}, \eqref{Young3} and \eqref{Young4}.
Thus we may identify the parameters as
\begin{align}
 \tilde \Lambda^{(1)} = \Lambda^{(1)} \, , \qquad
\tilde \Lambda^{(2)} = \Lambda^{(2)} \, , \qquad
\tilde m = m \, . 
\label{idyoung}
\end{align}
Notice that the condition for $\tilde m$ \eqref{codtm1} and $\eqref{codtm2}$
reproduces the selection rule in \eqref{selection}. {}From the ADM mass
of the geometry, we may read off the boundary conformal dimension from the
classical geometry as (see (3.16) of \cite{Castro:2011iw})
\begin{align}
 h =  - \frac{c}{6} M_N =  h (\Lambda^{(1)}, \omega ; \Lambda^{(2)} , m) - \frac{c}{24} \, ,
\end{align}
where we have used
\begin{align}
 C^{(a)} (\Lambda^{(a)}) = \frac12 \sum_i \left[ (l_i^{(a)} + \rho^{(a)}_i)^2 - ( \rho^{(a)}_i)^2 \right] 
\end{align}
for $a=1,2$.
Here $M_N$ is define in \eqref{M_N} and in terms of the holonomy matrix it is written as
\begin{align}
 M_N = - \frac{1}{4 \epsilon_N} \left( \sum_{l=1}^{N+1} \theta_l ^2 - \sum_{\bar \jmath =N+2}^{2N+1} \theta^2_{\bar \jmath} \right) \, .
\end{align}
In this way, we have seen that the boundary conformal dimension from the geometry
reproduces the one of dual CFT in \eqref{confcft}, where the shift $-c/24$ comes from
the change of worldsheet geometry from the cylinder of boundary AdS to the complex plane.

\subsection{$W$-algebra charges}
\label{wcharges}

In this subsection, we would like to examine the charges of $W$-algebras for states primary to 
the ${\cal N}=2$ super $W_{N+1}$ algebra and compare them to the charges for the smooth solutions
of the gravity theory.
The charges for the classical geometry have
been computed in the previous section as \eqref{chargeu1}, \eqref{charget+} and \eqref{charget-}.
In order to compute charges in the CFT side, 
we utilize the free field realization of the ${\cal N}=2$ super $W_{N+1}$ algebra 
 in \cite{Ito:1991wb} and construct states primary to the symmetry algebra in terms of free fields.
The descendants are then obtained by the action of $W$-algebra generators to the primary states, see \eqref{primarycd} and \eqref{descendants} below.
The charges of $W$-algebra can be read off from the action of zero modes of 
the $W$-algebra generators to these primary states. 
Review articles on $W$-algebra may be found in \cite{Bilal:1991eu,Bouwknegt:1992wg}.

First we focus on the NS-sector and then move to the R-sector.
We introduce the super coordinate $Z=(z,\theta)$ with a Grassmanian variable $\theta$ and the super
derivative $D = \partial_\theta + \theta \partial_z$. Furthermore, $2N$ superfields are written as
$\Phi ^j (Z) = \phi^j (z) + i \theta \psi^j (z)$ $(j=1,2,\ldots 2N)$ with operator products
\begin{align}
 \phi^i (z) \phi^j (0) \sim - \delta_{i,j} \ln z \, , \qquad
 \psi^i (z) \psi^j (0) \sim \delta_{i,j} z^{-1} \, .
 \label{freeope}
 \end{align}
With the preparation we introduce a Lax operator by \cite{Evans:1990qq,Komata:1990cb}
\begin{align}
 L(Z) 
       &= ( a_0 D + i \Theta_{2N+1} (Z) ) (a_0 D + i  \Theta_{2N} (Z) ) \cdots ( a_0 D + i \Theta_{1} (Z) ) 
       \label{laxop} \\ \nonumber 
       &= ( a_0 D)^{2N+1}  + \sum_{j=2}^{2N+1} U_{\frac{j}{2}} (Z) ( a_0 D )^{2N+1 - j} \, ,
\end{align}
where $U_{j/2} (Z) $ $(j=2,3,\ldots , 2N+1)$ are the generators of ${\cal N}=2$ super $W_{N+1}$ algebra.
Here $\Theta_j (Z) = (-1)^{j-1} (\lambda_j - \lambda_{j-1}) \cdot D \Phi (Z) $ $(\lambda_0 = \lambda_{2N+1} =0)$, and the normal ordering is implicitly assumed when operators are inserted at the same position.
Moreover, $\lambda_j$ is the fundamental weight of $\text{sl}(N+1|N)$, see appendix \ref{superalgebra}.
{}From this equation, we have
\begin{align}
 U_{\frac{j}{2}} (Z) = \zeta \sum_{ 1 \leq l_1 < \cdots < l_j \leq 2N+1} (-1)^{\sum_{p=1}^j l_p} 
 ( a_0 D + i \Theta_{l_j} ) \cdots (a_0 D + i  \Theta_{l_2} ) ( i \Theta_{l_1}) 
 \label{udef}
\end{align}
with $\zeta = - 1$ for $j =1,2$ mod 4 and $\zeta = + 1$ for  $j = 0,3$ mod 4.
We may redefine
\begin{align}
 U_{p-1} (Z)= J_{p-1} (z) + i \theta [ G^+_{p-\frac12} (z) + G^-_{p-\frac12} (z) ] \, , \quad
 U_{p-\frac12} (Z) = a_0 [ i G^-_{p-\frac12} (z) +   \theta T_p (z)] \, ,
 \label{Wgenerators}
\end{align}
then $\{ J_1 , G^\pm_{3/2} , T_2 \}$ generate the ${\cal N}=2$ superconformal algebra as a subalgebra.
For instance, $T  \equiv T_2 - \frac12 \partial J_1$ corresponds to the energy momentum tensor
with the central charge $c = 3N(1 - (N+1) a_0^2 )$. 
The parameter $a_0$ takes
\begin{align}
 a_0^2 = \frac{1}{N+k+1} \sim - \frac{c}{12 \epsilon_N} = - \hat k
\end{align}
for the $\mathbb{CP}^N$ model  \eqref{CPNmodel2}, and it is proportional
to $c$ in the limit we are interested in.
The other fields are not primary with respect to
the energy momentum tensor, and we need to modify the operators by using lower dimension operators.
See, e.g, \cite{Ozer:2001bi} for the explicit form of first few operators.

Let us consider a vertex operator
\begin{align}
 V_\Lambda (Z) = \exp ( i a_0 \Lambda \cdot \Phi (Z) ) \, ,
 \label{vo}
\end{align}
where 
$\Lambda$ takes a  weight of $\text{sl}(N+1|N)$ as
\begin{align}
 \Lambda = \sum_{l=1}^{2N} \Lambda_l \lambda_l 
\end{align}
with non-negative integer $\Lambda_l$.
As shown in \cite{Ito:1991wb} and appendix \ref{degenerate},
the corresponding states have the maximal number of fermionic null vectors.
The weight
 can be written in terms of bosonic subalgebra $\text{sl}(N+1) \oplus \text{sl}(N) \oplus \text{u}(1)$ 
through \eqref{super2boson}.
The $W$-algebra charges $u_{j/2}$ can be read off from the operator product expansions as
\begin{align}
& U_p (Z_1) V_\Lambda (Z_2) \sim u_p (\Lambda) V_\Lambda (Z_2) Z_{12}^{-p} + \cdots \, , \\
& U_{p-\frac12} (Z_1) V_\Lambda (Z_2) \sim u_{p-\frac12} (\Lambda) V_\Lambda (Z_2) \theta_{12} Z_{12}^{-p} + \cdots \, .
\end{align}
Here the dots denote less singular terms and 
$Z_{12} = z_1 - z_2 - \theta_1 \theta_2$ and $\theta_{12} = \theta_1 - \theta_2$. 
With the help of \eqref{udef} and 
\begin{align}
 (a_0 D_1 + i \Theta_j (Z_1)) V_\Lambda (Z_2) \sim  a_0 [ D_1 
 - (-1)^{j} (\lambda_j - \lambda_{j-1} ) \cdot \Lambda \theta_{12} Z_{12}^{-1} ] V_\Lambda (Z_2)  + \cdots \, ,
\end{align}
we can easily obtain the charges for $J_1$ and $T \equiv T_2 - \frac12 \partial J_1$ as
\begin{align}
 q = u_1 = - a_0^2 m \, , \qquad
 h =  a_0^{-1} u_{\frac{3}{2}} + \tfrac12 u_1 
 =  \frac{a_0^2}{2} ( \Lambda  + \rho ) \cdot \Lambda  \, .
 \label{lowu}
\end{align}
Here $\rho$ is the Weyl vector for $sl(N+1|N)$ as in \eqref{rho0}.
These formulas may differ from  \eqref{confcft0}
and \eqref{qcft0} for the Kazama-Suzuki model \eqref{CPNmodel2}
by integer numbers $n,n'$, which are actually irrelevant at large $c$ limit.
In other words, the states in the Kazama-Suzuki model may differ from states with the label $\Lambda$
in \eqref{vo}, but the difference would be relevant only when we consider the next order of $1/c$.%
\footnote{Some comments on this issue may be found in section \ref{conclusion} and appendix \ref{degenerate} below.} 
For other charges, it is tedious but straightforward computations to obtain. 
For instance, the charge $u_2$ is computed as
\begin{align}
 a_0^{-4} u_2 = - \tfrac{1}{2} \sum_{j=1}^{N} ( l_j^{(2)} + \rho^{(2)}_j )^2 + \frac{N-1}{2N} m^2 + \frac{N-1}{2} m + \frac{N(N^2 -1)}{24} \, .
 \label{utwo}
\end{align}

The comparison with the gravity results in \eqref{chargeu1}, \eqref{charget+} and \eqref{charget-}
is not possible yet since the definition of $W$-algebra generators is not the same between the 
gravity side and the CFT side. In particular, it is known that the naive definition of energy momentum
tensor from the gravity theory in \eqref{Wgenerators_g} does not includes the $\text{u}(1)$ part,
and we should take care of this fact as mentioned in \cite{Hanaki:2012yf} (see also \cite{Henneaux:1999ib}).
Thus we should reorganize the generators of $W$-algebra so as to be primary with respect to the
modified energy momentum tensor. For the first three terms, they are given as \cite{Hanaki:2012yf}%
\footnote{The relation to the notation in \cite{Hanaki:2012yf} is given by $J_1 = - J$,
$T_2 = - T$, $J_2 = W_2^-$ and $a_0^2 = - a^2$.}
\begin{align}
& \tilde J_1 = J_1 \, , \nonumber\\
& \tilde T_2 = T_2 - \frac{1}{2} \partial J_1 - \frac{J_1J_1}{2N(1+N)a_0^2} \, ,\\
& \tilde J_2 = J_2 + \frac{(1-N)a_0^2 \partial J_1}{2}  + \frac{(1-N) J_1J_1}{2N}   
  - \frac{(N - 1) a_0^2 \tilde T_2 }{3} \nonumber
\end{align}
for large $a_0^2$. On the other hand, the ${\cal N}=2$ duality  suggests that%
\footnote{We use the expression in (5.15) of \cite{Creutzig:2012xb} with $\lambda = N+ 1$.}
\begin{align}
 J_s \leftrightarrow - \frac{2N+1}{4 s - 2} V^{(s)+}_{s-1} + \frac{1}{2} V^{(s)-}_{s-1} \, , \qquad
 T_s \leftrightarrow V^{(s)+}_{s-1}
\end{align}
for $s \geq 2$. Therefore, we can see the match of the first three charges of $W$-algebra as
\begin{align}
&a_0 v^{(1)}_- = u_1 \,  , \nonumber \\
&  v^{(2)}_+ + v^{(2)}_-   =  h - \frac{(u_1)^2}{2N(N+1)a_0^2}
 -  \frac{c}{24}  \,  ,\\
 & a_0^2 v^{(2)}_ -  
  = u_{2}  - \frac{ (1-N)a_0^2 u_1}{2}  + \frac{(1-N) (u_1)^2}{2N}  \nonumber
\end{align} 
by properly choosing the relative normalizations.

Let us turn to the R-sector. In order to discuss this sector, it is useful to utilize
the spectral flow symmetry of ${\cal N}=2$ superconformal algebra  introduced in 
\cite{Schwimmer:1986mf}. The algebra is invariant under the following transformation as
\begin{align}
& J^ {\eta}_1(z) = J_1 (z) + \frac{c \eta}{3 z} \, , \nonumber \\
& G^{\pm, \eta }_{3/2} (z) = z^{\pm \eta} G^{\pm}_{3/2} (z) \,  , \\
& T^\eta_2 (z) = T_2 (z) + \eta J_1 (z) + \frac{c \eta^2}{6 z^2}\,  , \nonumber
\end{align}
where $\eta$ is a continuous parameter.
If we set $\eta = 1/2$, then the transformation maps the NS-sector to the R-sector.
It is argued that the spectral flow is generated by the operator \cite{Lerche:1989uy} 
\begin{align}
  {\cal U}_\eta (z) = \exp \left( - i \eta \sqrt{\frac{c}{3}} \varphi (z)  \right) \, , \qquad
  J_1 (z) = i \sqrt{\frac{c}{3}} \partial \varphi (z) \, .
\end{align}
However, the bosonic subsector of $W$-algebra generators defined in \eqref{Wgenerators_g} 
from the gravity side decouple with the $\text{u}(1)$ sector by definition.
Therefore, with the basis, the bosonic generators are invariant under the spectral flow,
and this implies that the charges in the R-sector match once  the
correspondence of charges is shown in the NS-sector.

\subsection{Null vectors v.s. higher spin symmetry}
\label{nullvs}

As mentioned above,  descendants in the theory based on the ${\cal N}=2$ super $W_{N+1}$-algebra
are obtained by the action of the $W$-algebra generators to the primary states.
Let us denote $|\Lambda \rangle$ as a primary state and the mode expansions of $W$-algebra generators 
as
\begin{align}
 W^{(s)a} (z) = \sum_r \frac{W^{(s)a}_r}{z^{r+h^{(s)a}}} ~,
\end{align}
where $W^{(s)0} (z) = J_s (z) $, $W^{(s)\pm} (z) = G^{\pm}_{s+1/2} (z) $, $W^{(s)1} (z) = T_{s+1} (z) $ and
 $h^{(s)a}$ denote their conformal weights. The sum runs over $r \in \mathbb{Z}$ for bosonic operators and fermionic 
operators in the R-sector. For fermionic operators in the NS-sector, it runs over $r \in \mathbb{Z}+1/2$.
The condition to be primary can be then written as
\begin{align}
 W^{(s)a}_{r} | \Lambda \rangle = 0 \, , \qquad r > 0 \, ,
 \label{primarycd}
\end{align}
and the descendant states are generated as
\begin{align}
| X \rangle =  W^{(s_1)a_1}_{-r_1} W^{(s_2)a_2}_{-r_2}  \cdots W^{(s_l)a_l}_{-r_l} |\Lambda \rangle 
\label{descendants}
\end{align}
with $r_i > 0 $. When a descendant satisfies the condition of primary, the state is null
and should be removed from the spectrum. 

These null states are constructed by the action of null vectors to primary states.
For our case, there are the maximally possible number of null vectors from each state
associated with the weight $\Lambda$ of $\text{sl}(N+1|N)$ Lie superalgebra as in \eqref{vo}.
In the NS-sector,  ``independent'' null vectors have been investigated in \cite{Ito:1991wb}
(see also appendix \ref{degenerate}),
and there are $2N$ fermionic null vectors
 at level $\Lambda_I + 1/2$ $(I=1,2,\ldots 2N)$.
It will be useful to notice that
\begin{align}
 \Lambda_{2j-1 } = r^{(1)}_{j} - r^{(2)}_{j} - \Lambda_{2N} \, , \qquad
 \Lambda_{2j - 2} = - r^{(1)}_j + r^{(2)}_{j-1} + \Lambda_{2N} 
\end{align}
with
\begin{align} 
 \Lambda_{2N} = \frac{|\Lambda^{(1)}|}{N+1} - \frac{|\Lambda^{(2)}|}{N} + \frac{m}{N(N+1)} 
 \label{lambda2N}
\end{align} 
in terms of the bosonic subalgebra.  
As an example, let us see what happens for the chiral primary states with $h = q/2$.
These states are labeled as \cite{Gepner:1988wi}
\begin{align}
 \Lambda^{(1)}_j = \Lambda^{(2)}_j ~ (j = 1,2, \ldots , N-1) \, , \qquad
  m = \sum_{j=1}^N j \Lambda^{(1)}_j \, ,
\end{align}
which implies
\begin{align}
 \Lambda_{2j - 1} =0
\end{align}
for all $j$.
Therefore, we have $N$ independent null vectors at level $1/2$.
One of them is given by the action of $(G_{3/2}^+)_{-1/2}$, which arises from the definition of chiral primary.
The others are generated by $N-1$ independent linear combinations of $(G^+_{s+1/2})_{-1/2}$ with
$s=1,2,\ldots, N$.

It was pointed out in  \cite{Perlmutter:2012ds} that these null vectors should be identified as 
the residual higher spin symmetries of the smooth gauge field configuration dual to the primary state. 
When we perform the path integral of the gauge theory, we have to divide the directions of 
gauge symmetry, which corresponds to removing the null vectors.
For both cases with anti-periodic and periodic Killing spinors, 
we can read off from \eqref{spinorphase} that
 the Killing spinors corresponding to fermionic higher spin symmetry have 
 $w$-dependence
 as $\exp ( - (\theta_l - \theta_{\bar \jmath} ) w )$.  If we consider the spinors associated with
$\epsilon^{\bar \imath , j}$ in \eqref{killingspinor}, then we have similarly the Killing spinors
behaving as $\exp (  (\theta_j - \theta_{\bar \imath} ) w )$.
Utilizing the parameters in \eqref{Young1}, \eqref{Young2} and \eqref{ntilde}, we see that 
\begin{align}
 -  i (\theta_l - \theta_{N+1+j}) 
  = \tilde n^{(1)}_i  - \tilde n^{(2)}_j  - \frac{\tilde m}{N(N+1)} \, .
\end{align}
Therefore, the negative mode number of the Killing spinor  with $\epsilon^{j , N + 1 + j }$ or $ \epsilon^{ N+1+ j +1 , j }$ coincides with the level of fermionic 
null vector $\Lambda_I + 1/2$.
The other negative mode numbers can be obtained by the shifted Weyl reflections of $\text{sl}(N+1|N)$,
which are the same as those of bosonic subalgebra $\text{sl} (N+1) \oplus \text{sl} (N)$.
These additional Killing spinors should correspond to null vectors appearing as descendants of  the
independent null vectors as in the bosonic case \cite{Niedermaier:1991cu,Perlmutter:2012ds}.%
\footnote{In the bosonic $W_N$ minimal model, the Weyl invariance of the null vector structure was shown in 
\cite{Niedermaier:1991cu}.  In our supersymmetric model, it is an open problem to proof (or disproof) this.}
The maximally supersymmetric geometry also preserves the maximal number of bosonic higher spin symmetry.
This is because the bosonic subgroup of the holonomy matrix along the spatial cycle is given 
by the center of $\text{SL}(N+1) \otimes \text{SL}(N)$ up to a similarity transformation,
see \eqref{holNS} and \eqref{holR} above.
In particular, we can show as in  \cite{Perlmutter:2012ds} that the  $w$-dependence of the Killing vector
 is $\exp ( - i (\tilde n^{(a)}_i - \tilde n^{(a)}_j ) w ) $ for $a =1,2$.
Therefore the negative mode number of the higher spin bosonic symmetry
is given by $(n^{(a)}_i - n^{(a)}_j)$ with $i < j$.
These Killing vectors should correspond to bosonic null vectors appearing as descendants of fermionic 
independent null vectors.

{}From the relation between higher spin symmetry and null vectors, we can say that the 
geometry dual to the primary states should have the maximal number 
of higher spin symmetry. Moreover, we may obtain the one-loop partition function of the
gravity theory from the relation to the CFT. The one-loop partition function of the CFT 
can be written as a sum of characters of representation $\Xi = (\Lambda^{(1)} ; \Lambda^{(2)}, m)$ as
\begin{align}
  Z _\text{1-loop}^\text{CFT} (q)  = \sum_\Xi |\text{ch}_\Xi^\text{NS,R} \, (q) |^2 \, , \qquad
  \text{ch}_\Xi ^\text{NS,R} \, (q) = \text{tr}_\Xi \,  q^{L_0} \, .
	\label{CFTpf1}
\end{align}
Here the trace is over the states obtained by the action of $W$-algebra generators to the primary
state with label $\Xi$ modulo the null vectors in the NS-sector or in the R-sector. 
{}From the CFT expression in \eqref{CFTpf1} we expect that the one-loop partition function of
the gravity theory can be obtained following \cite{Castro:2011iw}
as the sum over the contributions from each smooth geometry.
See section \ref{conclusion} for some discussions.

\section{Conclusion and discussions}
\label{conclusion}

In this paper, we have studied purely bosonic conical defects in $\text{sl}(N+1|N) \oplus \text{sl}(N+1|N)$
Chern-Simons gauge theory with the maximal number of fermionic higher spin symmetry,
where both anti-periodic and periodic boundary conditions of Killing spinors can be chosen.
The gauge field configuration then is parametrized by two sets of integer number.
The holonomy matrix of the gauge field configuration is given by a center of 
bosonic subgroup up to a similarity transformation, which implies that the conical defect
can be mapped to a non-singular geometry. As in \cite{Castro:2011iw} we have extended
the class of smooth geometry, which is now labeled by two Young diagrams 
$\tilde \Lambda^{(a)}$ with $a=1,2$ and an integer $\tilde m$.
The smooth geometry is proposed to be dual to a primary state in the CFT with
${\cal N}=2$ $W_{N+1}$ symmetry at the limit of large central charge $c \to \infty$.
The primary state is labeled by $\text{sl}(N+1|N)$ weight as in \eqref{vo},
which can be expressed by two Young diagrams 
$\Lambda^{(a)}$ and $\text{u}(1)$ charge $m$.
We identify the labels as $\tilde \Lambda^{(a)}=\Lambda^{(a)}$ and $\tilde m = m$.
Moreover, the cases with anti-periodic and  periodic Killing spinors are mapped to
NSNS-sector and RR-sector of the CFT, respectively.
This proposal is checked by comparing some $W$-algebra charges.
The null vectors in the CFT are identified as the residual higher spin symmetry of the smooth geometry.

Once we know the relation between the null vectors and the residual higher spin symmetry,
we can guess the gravity partition function by following 
\cite{Perlmutter:2012ds}.
For the bosonic part of the gravity partition function, we can just use the result of
\cite{Perlmutter:2012ds} as
\begin{align}
 Z^\text{B}_\text{1-loop} (\Xi) = \frac{ \prod_{1 \leq i < j \leq N+1} |1 - q^{\tilde n^{(1)}_i - \tilde n^{(1)}_j} |^2  
 \prod_{1 \leq i < j \leq N}  |1 - q^{\tilde n^{(2)}_i - \tilde n^{(2)}_j} |^2 }{ \prod_{n=1}^\infty |1 - q^n|^{4N}} \,  ,
\end{align}
where we have also included the contribution from the $\text{u}(1)$ part.
{}For the fermionic part, we may have for the NSNS-sector as
\begin{align}
 Z^\text{NS}_\text{1-loop} (\Xi) = \frac{ \prod_{n=1}^\infty |1 + q^{n-1/2}|^{4N}}
  { \prod_{1 \leq i  \leq N+1 } \prod_{1 \leq  j \leq N} 
   |1 + q^{|\tilde n^{(1)}_i - \tilde n^{(2)}_j - \tilde m /N(N+1)|} | ^2 } \, ,
\end{align}
and for the RR-sector as
\begin{align}
 Z^\text{R}_\text{1-loop} (\Xi) = \frac{ \prod_{n=1}^\infty |1 + q^{n}|^{4N}}
  { \prod_{1 \leq i  \leq N+1 } \prod_{1 \leq j \leq N} 
   |1 + q^{|\tilde n^{(1)}_i - \tilde n^{(2)}_j - \tilde m /N(N+1)|} | ^2} \, .
\end{align}
The one-loop partition function of the gravity theory is obtained by the product of bosonic 
and fermionic contributions as
\begin{align}
 Z_\text{1-loop} (\Xi) = Z^\text{B}_\text{1-loop}(\Xi) Z^\text{NS,R}_\text{1-loop} (\Xi) \, .
\end{align}
For example,  the AdS background corresponds to the choice
$\tilde n^{(a)}_j =\rho^{(a)}_j$ and $\tilde m = 0$ in the NSNS-sector, which leads to
\begin{align}
 Z_\text{1-loop} (0)  = Z^\text{B}_\text{1-loop}(0) Z^\text{NS}_\text{1-loop} (0) 
  = \prod_{s=1}^N \prod_{n=s} ^\infty \left | \frac{1 + q^{n+\frac12}}{1 - q^n} \right | ^2
   \prod_{s=2}^{N+1} \prod_{n=s} ^\infty \left | \frac{1 + q^{n-\frac12}}{1 - q^n} \right | ^2 \, .
\end{align}
This is actually the same as the vacuum character of ${\cal N}=(2,2)$ super $W_{N+1}$ algebra.
The gravity partition function may differ from the CFT one by an overall factor
as in the bosonic case, see the end of section 6 in \cite{Perlmutter:2012ds}.
It is an important open problem to reproduce the above expression by directly computing
one-loop determinants of the supergravity.
Moreover, we should compare the gravity partition function with the CFT one,
which may be possible by utilizing the expression in \cite{Candu:2012jq} or by generalizing
the null vector analysis in \cite{Niedermaier:1991cu} to our supersymmetric case.
In particular, it is interesting to understand the structure of null vectors 
in the R-sector along with the detailed analysis on the NS-sector.

One of the motivation to study this semi-classical limit of the ${\cal N}=2$ duality is that we could have a AdS/CFT
correspondence involving purely three dimensional Chern-Simons theory without
any matter fields coupled unlike for the bosonic case. Without the matter fields, we
may have a chance to proof the duality by the application of the Drinfeld-Sokolov reduction.
For further understanding, 
we would like to study the duality beyond
the large $c$ limit by examining $1/c$ corrections.
For instance, at the next order of $1/c$, we can see the difference between the states in the 
Kazama-Suzuki model \eqref{CPNmodel} and the states corresponding to the vertex operator 
\eqref{vo}, see appendix \ref{degenerate}. {}From the experience on the bosonic case in \cite{Castro:2011iw}, it is natural
to identify the states with $\hat \Lambda = 0$ in \eqref{lambdap} as the smooth geometry,
and those with $\hat \Lambda \neq 0$ as a geometry dressed by perturbative corrections.
Since the Kazama-Suzuki model is an unitary model with ${\cal N}=2$  $W_{N+1}$ symmetry,
it would be important to see what kind of corrections make the theory unitary.
It is also worth to study conical defects for so$(N)$ holography 
\cite{Ahn:2011pv,Gaberdiel:2011nt,Candu:2012ne} and  for
${\cal N}=1$ holography \cite{Creutzig:2012ar}.
In particular, it was argued in \cite{Candu:2012ne} that the finite $N$ effects
for the so$(N)$ holography are a bit more complex than the su$(N)$ case.
It would be also interesting to study black hole solutions in the
higher spin supergravity and see the relation to the duality.
See a review \cite{Ammon:2012wc} for the bosonic case.
Some higher spin black holes in the supergravity 
have been already constructed in \cite{Tan:2012xi,Datta:2012km}.

\subsection*{Acknowledgements}

We are grateful to T.~Creutzig, R.~Gopakumar, P.~R{\o}nne and T.~Ugajin for useful discussions.
The work of YH was supported in part by Grant-in-Aid for Young Scientists (B) from JSPS.

\appendix

\section{$\text{sl}(N+1|N) $ Lie superalgebra}
\label{superalgebra}

We summarize here useful formulas on $\text{sl}(N+1|N) $ Lie superalgebra.

\subsection{Generators, roots and weights}

The generators of $\text{sl}(N+1|N) $ Lie superalgebra can be described 
$( (N+1)+ N )  \times ((N+1) + N )   $ supermatrices of the form
\begin{align}
 M= \left(
 \begin{array}{cc}
  A & B \\
  C & D
 \end{array}  
  \right) \,  , \qquad \text{str} \, M = \text{tr} \, A - \text{tr} \, D = 0 \, ,
\end{align}
where $A,D$ are even elements and $B,C$ are odd elements with respect to the
$\mathbb{Z}_2$ grading of superalgebra. The traceless part of $A,D$ generate
$\text{sl}(N+1), \text{sl}(N)$ bosonic subalgebra, respectively, and the centralizer of
the bosonic subalgebra is $\text{u}(1)$. The bosonic subalgebra is thus
$\text{sl}(N+1) \oplus \text{sl}(N) \oplus \text{u}(1)$.

The $\text{sl}(N+1|N) $ Lie superalgebra has a special property that 
we can choose a completely odd simple root system.%
\footnote{Such Lie superalgebras are given by
$\text{sl}(N\pm 1 | N)$, $\text{osp}(2N \pm 1 | 2N)$, $\text{osp}(2N|2N)$, $\text{osp}(2N+2|2N)$ and  $\text{D}(2,1;\alpha)$
with $\alpha \neq 0, \pm 1$, see, e.g., \cite{Frappat:1996pb}.
} 
We introduce two orthogonal bases $\varepsilon_i$ $(i=1,2,\ldots,N+1)$ and 
$\delta_i$ $(i=1,2,\ldots,N)$, which satisfy%
\footnote{We borrow the notations in \cite{Ito:1991wb}.}
\begin{align}
 \varepsilon_i \cdot \varepsilon_j = \delta_{i,j} \, , \qquad
 \delta_i \cdot \delta_j = - \delta_{i,j} \, .
\end{align}
Then the odd simple roots can be expressed as
\begin{align}
 \alpha_{2i - 1} = \varepsilon_i - \delta_i \, , \qquad \alpha_{2i} = \delta_i - \varepsilon_{i+1}
\end{align}
for $i=1,2, \ldots , N$. Then positive roots are 
\begin{align}
  \alpha_i + \alpha_{i+1} + \cdots + \alpha_j 
\end{align}
with $i \leq j$, where the root is even (odd) when $i - j $ is even (odd).
The fundamental weights are defined by
\begin{align}
 \alpha_i \cdot \lambda_j = \delta_{i,j} \, ,
\end{align}
which can be expressed by the simple roots as
\begin{align}
 \lambda_{2i} = \alpha_1 + \alpha_3 + \cdots + \alpha_{2i - 1} \, , \qquad
 \lambda_{2i-1} = \alpha_{2i} + \alpha_{2i+2} + \cdots + \alpha_{2N} \, . 
\end{align}

As mentioned above, the $\text{sl}(N+1|N)$ superalgebra has $\text{sl}(N+1) \oplus \text{sl}(N) \oplus \text{u}(1)$
bosonic subalgebra. The abelian factor $\text{u}(1)$ is generated by
\begin{align}
  \nu = \sum_{i=1}^N (\lambda_{2i} - \lambda_{2i-1}) \, ,
  \label{abelian}
\end{align} 
whose norm is $\nu \cdot \nu = - N(N+1)$.
The simple roots $\alpha^{(1)}_i$ and the fundamental weights $\lambda^{(1)}_i$ 
for $\text{sl}(N+1)$ subalgebra are 
\begin{align}
&\alpha^{(1)}_i = \alpha_{2i-1} + \alpha_{2i} = \varepsilon_i - \varepsilon_{i+1} \, , \\
&\lambda^{(1)}_i = \sum_{j=1}^{2i-1} (-1)^{j-1} \lambda_j + \frac{i}{N+1} \nu 
                      = \sum_{j=1}^i \varepsilon_j - \frac{i}{N+1} \sum_{j=1}^{N+1} \varepsilon_j 
\end{align}
with $i=1,2,\ldots , N$.  In the same way, the simple roots $\alpha^{(2)}_i$ and the fundamental weights $\lambda^{(2)}_i$ 
for $\text{sl}(N)$ subalgebra are 
\begin{align} 
&\alpha^{(2)}_i = \alpha_{2i} + \alpha_{2i+1} = \delta_i - \delta_{i+1} \, , \\
&\lambda^{(2)}_i = \sum_{j=1}^{2i} (-1)^{j} \lambda_j - \frac{i}{N} \nu 
                      = - \sum_{j=1}^i \delta_j + \frac{i}{N} \sum_{j=1}^{N} \delta_j 
\end{align}
with $i=1,2,\ldots , N-1$. The Weyl vector is
\begin{align}
 \rho = \sum_{i=1}^{2N} \lambda_i = 2 ( \rho^{(1)} + \rho^{(2)} ) \, ,
 \label{rho0}
\end{align}
where the Weyl vectors $\rho^{(1)},\rho^{(2)}$ for $\text{sl}(N+1),\text{sl}(N)$ are
\begin{align}\label{rho1}
 &\rho^{(1)} = \sum_{i=1}^{N} \lambda^{(1)}_i = 
  \sum_{i=1}^{N+1} \rho^{(1)}_i \varepsilon_i  = \sum_{i=1}^{N+1} \left( \tfrac{N+2}{2} - i\right) \varepsilon_i \, ,\\
 &\rho^{(2)} = \sum_{i=1}^{N-1} \lambda^{(2)}_i 
   = - \sum_{i=1}^{N} \rho^{(2)}_i \delta_i = - \sum_{i=1}^{N} \left( \tfrac{N+1}{2} - i\right) \delta_i \, .
    \label{rho2}
\end{align}

A weight for $\text{sl}(N+1|N)$ Lie superalgebra is expressed as
\begin{align}
 \Lambda =  \sum_{i=1}^{2N} \Lambda_i \lambda_i 
 \label{hwsl}
\end{align}
with non-negative integers $\Lambda_i$. In terms of bosonic subalgebra,
the weight may be expressed as
\begin{align}
 \Lambda  = \sum_{i=1}^{N} \Lambda^{(1)}_i \lambda^{(1)}_i
  + \sum_{i=1}^{N-1} \Lambda^{(2)}_i \lambda^{(2)}_i + \frac{m}{N(N+1)}\nu 
  \label{super2boson}
\end{align}
with
\begin{align}
 \Lambda^{(1)}_i = \Lambda_{2i} + \Lambda_{2i-1} \, ,\qquad
 \Lambda^{(2)}_i = \Lambda_{2i} + \Lambda_{2i+1}
\end{align}
and
\begin{align} 
  m =   \sum_{i=1}^N (i \Lambda_{2i} - (N+1-i) \Lambda_{2i-1}) \, .
\end{align}
With the decomposition, the weight may be labeled by two Young diagrams 
corresponding to $\Lambda^{(a)} = \sum_j \Lambda^{(a)}_j \lambda^{(a)}_j$ with $a=1,2$
along with an integer $m$.
The diagrams have $r^{(a)}_j$ boxes in the $j$-th low with 
\begin{align}
 r^{(1)}_j = \sum_{i=j}^{N} \Lambda^{(1)}_j \, , \qquad
 r^{(2)}_j = \sum_{i=j}^{N-1} \Lambda^{(2)}_j \, .
\end{align}
In the orthogonal basis, the weights are decomposed as
\begin{align}
 \Lambda^{(1)} = \sum_{j=1}^{N+1} l^{(1)}_j \varepsilon_j \, , \qquad
 \Lambda^{(2)} = - \sum_{j=1}^{N} l^{(2)}_j \delta_j
\end{align}
with
\begin{align}
 l^{(1)}_j = r^{(1)}_j - \frac{|\Lambda^{(1)}|}{N+1} \, , \qquad
 l^{(2)}_j = r^{(2)}_j - \frac{|\Lambda^{(2)}|}{N} \, .
 \label{l12}
\end{align}
Here $r^{(1)}_{N+1} = r^{(2)}_{N} = 0$ and $|\Lambda^{(a)}|$ is the number of boxes in the 
corresponding Young diagram.

\subsection{Generators}

The generators of $\text{sl}(N+1|N)$ are given by
\begin{align}
 V^{(s)+}_n ~ (s=2,3,\ldots, N+1) \, , \quad V^{(s)-}_n ~ (s=1,2,\ldots , N) \, , \quad F^{(s)\pm}_r ~
 (s=1,2,\ldots , N)
\end{align}
with $|n| \leq s-1, |r| \leq s-1/2$. We have utilized the principal embedding of $\text{osp}(1|2)$ 
into $\text{sl}(N+1|N)$ superalgebra, see \cite{Frappat:1996pb} for instance.
The embedded $\text{osp}(1|2)$ corresponds to $L_n = V_{n}^{(2)+}$ and $ G_r = F_r^{(1)+}$, 
which satisfy
\begin{align}
 [L_m , L_n ] = (m-n) L_{m+n}  \, , \quad
 [L_m , G_r] = (\tfrac12 m - r) G_{m+r} \, , \quad
 \{ G_r , G_s \} = 2 L_{r+s} \, .
\end{align} 
The (anti-)commutation relations to other generators are
\begin{align}
 & [ L_m , V^{(s)\pm}_n ] = (- n + m (s-1)) V^{(s)\pm}_{m+n} \, , \quad
 [ L_m , F^{(s)\pm}_r] = ( - r + m (s - \tfrac12)) F^{(s)\pm}_{m+n} \, , \nonumber \\ 
 & [G_{1/2} , V^{(s) +}_m] = - \tfrac{1}{2}(m -s + 1) F^{(s-1)+}_{m+1/2} \, , \quad 
 [G_{1/2} , V^{(s)-}_m] = - 2 F^{(s)-}_{m+1/2} \, , \\
 & \{ G_{1/2} , F^{(s-1)+}_r \} = 2 V^{(s)+}_{r+1/2} \, , \quad
 \{ G_{1/2} , F^{(s)-}_r \} = \tfrac{1}{2} (r - s + \tfrac12) V^{(s)-}_{r+1/2} \, . \nonumber 
\end{align}
Other (anti-)commutation relations can be found in \cite{Fradkin:1990qk}.

It might be useful to express the generators $L_m$ of sl(2) subalgebra in terms of supermatrix. 
We use 
\begin{align}
 L_n = \left(
 \begin{array}{c|c}
 K_n^{N+1} & 0 \\ \hline
  0 & K_n^{N}
 \end{array}  
 \label{L_n}
 \right) \, ,
\end{align} 
with
\begin{align}
 K_0^{M} = \frac12  \left(
  \begin{array}{ccccc}
  M -1 & 0& & & \\
 0 & M - 3 &  0& &  \\
  & & \ddots & &  \\
  & & 0& 3 - M & 0\\
& & & 0&  1- M 
 \end{array}   
 \right) \, , \label{K_0}
\end{align}
\begin{align} 
 K_{1}^{M} = -  \left(
  \begin{array}{cccccccc}
  0 & & & &  & & & \\
  \sqrt{M-1}& 0&  & & & & & \\
  0 & \sqrt{2(M-2)}& 0  & &  & &  &\\
  & & \ddots  &  & & &  &\\
  & &0 &  \sqrt{|i (M-i)|} & 0 & & & \\ 
  & & &  & \ddots & & & \\
  & & & &  & 0& \sqrt{M-1} & 0
 \end{array}  
 \right) \, , 
\end{align} 
\begin{align} 
 K_{-1}^{M} =   \left(
  \begin{array}{ccccccc}
  0&  \sqrt{M-1}& 0 & & &  &\\
  & 0  & \sqrt{2(M-2)}&   0 & &   &\\
   & &   &\ddots   &  & & \\
  & &   & 0 &\sqrt{|i (M-i)|} & 0 & \\ 
  & &   &  & & \ddots & \\
 & &  &  & & 0& \sqrt{M-1} \\
 & &  &  & & & 0
 \end{array}  
 \right) \, .
\end{align}
In particular, we find
\begin{align}
 \epsilon_N = \text{str}\, (L_0 L_0 ) = \text{tr} \, (K_0^{N+1} K_0^{N+1}) 
 - \text{tr} \, (K_0^{N} K_0^{N}) = \frac{N(N+1)}{4} \, .
\end{align}
For $J_o = V^{(1)-}_0$, we use
\begin{align}
 J_0 = \left(
 \begin{array}{c|c} 
 N \mathbf{1}_{(N+1) \times (N+1)} & 0\\ \hline
  0 & (N+1) \mathbf{1}_{N \times N}
 \end{array}  
 \right) \, ,
\end{align} 
where $\mathbf{1}_{M \times M}$ is the $M \times M$ identity matrix.
The normalization is 
\begin{align}
 t^{(s)}_- = \text{str} ( J_0 J_0 ) = - N (N+1) \, .
\end{align}

\section{Degenerate representations of ${\cal N}=2$ $W_{N+1}$ algebra}
\label{degenerate}

In this appendix we review the analysis in section 3.3 of \cite{Ito:1991wb} on degenerate 
representations of ${\cal N}=2$ $W_{N+1}$ algebra and their null vector structures.
We introduce $2N$ free bosons $\phi^j$ and fermions $\psi^j$ $(j=1,2,\ldots , 2N)$
with the operator products in \eqref{freeope}. We consider the primary fields of the
form
\begin{align}
 V_\Lambda (z) = \exp (i a_0 \Lambda \cdot \phi (z) ) \, .
\end{align} 
Here we have used
\begin{align}
 \Lambda = \sum_{j=1}^{2N} \Lambda_j \lambda_j
 \label{slNN+1}
\end{align}
in terms of $\text{sl}(N|N+1)$ fundamental weights $\lambda_l$.
At this stage $\Lambda_l$ takes any real number.  
We may express it in terms of bosonic subalgebras as
\begin{align}
 \Lambda  = \sum_{i=1}^{N} \Lambda^{(1)}_i \lambda^{(1)}_i
  + \sum_{i=1}^{N-1} \Lambda^{(2)}_i \lambda^{(2)}_i + \frac{m}{N(N+1)}\nu \, .
  \label{bosonicsub}
\end{align}
Descendants are generated by the action of negative modes of ${\cal N}=2$
$W_{N+1}$ currents as in \eqref{descendants}.  As explained in section \ref{nullvs}, some descendants
may satisfy the condition for primary fields \eqref{primarycd}. In that case, we can remove the
corresponding states from the spectrum in a consistent way, and this could happen
only  for restricted classes of $\Lambda$.

Before going into  detailed analysis on these degenerate representations,
we remark on a global symmetry. The charges of the $W$-algebra
for the primary operators can be computed as in section \ref{wcharges}.
Primary fields with different $\Lambda$ can have same $W$-charges,
and the corresponding states should be identified. As obtained in section 3.1 of \cite{Ito:1991wb} (for the bosonic case, see, 
e.g., \cite{Bilal:1991eu}),
the condition of the identification is 
\begin{align}
 - m + h_{2m} \cdot \Lambda = - m ' + h_{2m '} \cdot \Lambda \, , \qquad
 -n + h_{2n+1} \cdot \Lambda = - n ' + h_{2 n ' + 1} \cdot \Lambda 
\end{align}
where $m' , n'$ are obtained by permutations of $m,n$.
Here we have defined
\begin{align}
& h_{2m} = \lambda_{2m-1} - \lambda_{2m} = \lambda^{(2)}_{m-1} - \lambda^{(2)}_{m} - \frac{\nu}{N} \, , \\
& h_{2m+1} = \lambda_{2m +1} - \lambda_{2m} = \lambda^{(1)}_{m+1} - \lambda^{(1)}_{m} - \frac{\nu}{N+1} \, .
\nonumber
\end{align}
In particular, the states with $\Lambda$ and $-\rho - \Lambda$ are dual to each other.
In the following, we study the condition that null vectors appear, however we should 
remember that there are identifications among states as above.

In order to construct null fields, we utilize screening charges which commute with the 
${\cal N}=2$ $W_{N+1}$ generators. There are three types of screening charges. One
of them is obtained by the Hamiltonian reduction of $\text{sl}(N+1|N)$ WZW models
as \cite{Ito:1990ac}
\begin{align}
 S_j (z) = \alpha_j \cdot \psi e^{i a_0^{-1} \alpha_j \cdot \phi (z)}  \, .
 \label{screenf}
\end{align}
The other two are related to the bosonic subalgebras as \cite{Ito:1991wb} 
\begin{align}
 S^{(1)}_i = \left[ (\alpha_{2i} - \alpha_{2i-1} ) \cdot \partial \phi - 2 i a_0 (\alpha_{2i} \cdot \psi) (\alpha_{2i-1 } \cdot \psi)\right] e^{- i a_0 \alpha^{(1)} \phi}
 \label{screenb1}
\end{align}
for $i=1,2,\ldots , N$ and
\begin{align}
 S^{(2)}_i = \left[ (\alpha_{2i} - \alpha_{2i +1} ) \cdot \partial \phi + 2 i a_0 (\alpha_{2i} \cdot \psi) (\alpha_{2i +1 } \cdot \psi)\right] e^{ i a_0 \alpha^{(2)} \phi}
 \label{screenb2}
\end{align}
for $i=1,2,\ldots , N-1$.
Since the screening charges commute with the ${\cal N}=2$ $W_{N+1}$ generators,
we can construct a null field from a primary field $V_{\Lambda '}(z)$ as
\begin{align}
 \chi_\Lambda (z) = \int du_1 \cdots du_r S(u_1) \cdots S(u_r) V_{\Lambda '} (z) 
\end{align} 
if the integral exists  non-trivially. The integral contours are taken as in, e.g., \cite{KM}.

Let us start from a fermionic null field. Utilizing a fermionic screening operator \eqref{screenf}, we have
\begin{align}
 \chi_\Lambda (z) = \int d u S_j (u) V_{\Lambda '} (z) 
  = \int du (u - z)^{\alpha_j \cdot \Lambda '} \alpha_j \cdot \psi e^{i a_0 (\Lambda ' + \alpha_j a_0^{-2}) \phi  (z)} \, .
\end{align}
The integral exists for
\begin{align}
 1 + \alpha_j \cdot \Lambda ' = - N_j
\end{align}
with a non-negative integer $N_j$. Setting $\Lambda ' = - \rho - \Lambda - \alpha_j a_0^{-2}$,
we see that a null vector appears at the level $N_j + 1/2$ when 
\begin{align}
\Lambda_j = N_j
\end{align}
 in \eqref{slNN+1}.
 We have thus $2N$ independent conditions for the fermionic null vectors,
and maximally degenerate representations are given from
\begin{align}
 \Lambda = \sum_{j=1}^{2N} N_j \lambda_j 
\end{align}
with $N_j \geq 0$.
In terms of  bosonic subalgebras \eqref{bosonicsub}, we find the left hand side of 
\eqref{lambda2N} should be an integer number. This condition should be related to the
selection rule \eqref{selection} in the $\mathbb{CP}^N$ Kazama-Suzuki model \eqref{CPNmodel}.

We can construct a bosonic null field with a bosonic screening charge \eqref{screenb1}.
A null field is given by
\begin{align}
\chi_\Lambda (z) = \int d u_1 \cdots d u_{r_i} \prod_{j=1}^{r_i} S^{(1)}_i (u_j) V_{\Lambda '} (z)
\end{align}
when the integral leads to a non-zero value. The condition is
\begin{align}
 r_i - r_i + r_i (r_i - 1) a_0^2 - r_i a_0^2 \alpha^{(1)} \cdot \Lambda ' = - r_i s_i 
\end{align}
with positive integers $r_i,s_i$.%
\footnote{Contrary to the fermionic null fields, the states $\chi_\Lambda (z)$ are not
descendants for $r_i = 0$ nor $s_i=0$.}
We set $\Lambda ' = - \rho - \Lambda + r_i \alpha^{(1)}_i$.
Then a bosonic null vector appears at the level $r_i s_i$ for the case with 
\begin{align}
\Lambda^{(1)}_i = (r_i - 1) - s_i a_0^{-2}
\end{align}
 in \eqref{bosonicsub}.
In the similar manner, we can construct a null field using the other type of
screening charge \eqref{screenb2}. We find that a bosonic null vector
appears at the level $r_i ' s_i ' $ with positive 
integers $r_i' , s_i'$  for the case with 
\begin{align}
\Lambda^{(2)}_i = (r_i ' - 1) + s_i ' a_0^{-2}
\end{align}
in \eqref{bosonicsub}.

Notice that even if we want assign the maximal number of the bosonic null vector
conditions, we can do so only for $2N-1$ of them. In other words, we do not have 
condition for $m$ as in the selection rule \eqref{selection}. 
Moreover, the vacuum representation with $\Lambda = 0$ does not have bosonic null vectors of these kinds contrary to the bosonic case.
The representation with
 $N_j \in \mathbb{Z}$ (including $N_j = 0$) for all $j$ has independent fermionic null vectors 
as mentioned above, and bosonic null vectors may be generated by these independent ones.
In order to relate to the $\mathbb{CP}^N$ Kazama-Suzuki model \eqref{CPNmodel}, we should set 
$a_0^{-2} = k + N + 1$, which is an integer number. Therefore, in that case,
we could have the both types of null vectors simultaneously. 
In fact, as pointed out in \cite{Candu:2013uya}, the states in the $\mathbb{CP}^N$  Kazama-Suzuki model \eqref{CPNmodel}
may correspond to states labeled by
\begin{align}
 \Lambda = \sum_{j=1}^{2N} (\Lambda_j + a_0^{-2} \hat \Lambda_j ) \lambda_j \, .
 \label{lambdap}
\end{align}
Here we may need to set some of $\hat \Lambda_j $ non zero in order to explain the integers in 
\eqref{confcft0} and \eqref{qcft0}. However, in our case with $a_0^{-2} \to 0$, 
we cannot see $\hat \Lambda_j $ dependence in the level of independent fermionic null vectors.
{}From the degenerate representations obtained above, we may be able to construct unitary models,
 like the $\mathbb{CP}^N$ Kazama-Suzuki model \eqref{CPNmodel}, though it seems to be a quite
difficult problem. In particular, we have to find out fusion rules which are consistent with 
the positive norm spectrum. For the bosonic case, see \cite{Bilal:1991eu} for instance. 
It would be quite important to investigate on this issue.

\end{document}